\documentclass[aps,pra,twocolumn,bibnotes,,superscriptaddress]{revtex4-2}

\usepackage{amsmath}
\usepackage{amsthm}
\usepackage{latexsym}
\usepackage{amssymb}
\usepackage{bm}
\usepackage{bbm}
\usepackage{appendix}
\usepackage{epstopdf}
\usepackage{graphicx} 
\usepackage{physics}
\usepackage{color}
\usepackage{ragged2e}
\usepackage[colorlinks=true,linkcolor=blue,citecolor=green,plainpages=false,pdfpagelabels]{hyperref}
\usepackage[normalem]{ulem}
\usepackage{newlfont}
\usepackage{amsfonts}
\usepackage{amsthm}
\usepackage{float} 
\usepackage{ragged2e}

\usepackage[utf8]{inputenc}

\usepackage{amsfonts}
\usepackage{array}

\usepackage{geometry}
\usepackage{enumitem}

\usepackage{booktabs}
\usepackage{array}
\usepackage{graphicx}
\usepackage{epsfig}
\usepackage{mathrsfs}
\usepackage[thinc]{esdiff}
\usepackage{caption}
\usepackage{subcaption}

\usepackage{times}

\def\tr{\operatorname{tr}}

\DeclareOldFontCommand{\rm}{\normalfont\rmfamily}{\mathrm}

\begin{document}

\author{Brij Mohan}
\email{brijhcu@gmail.com}
\affiliation{Nano and Molecular Systems Research Unit, University of Oulu, 90014 Oulu, Finland}
\affiliation{Department of Physical Sciences, Indian Institute of Science Education and Research (IISER), Mohali, Punjab 140306, India}

\author{Tanmoy Pandit}
\email{tanmoy.pandit@vtt.fi }

\affiliation{VTT Technical Research Centre of Finland, Tietotie~3, Espoo, Finland}
\affiliation{QMill Oy Keilaranta 12 D, 02150 Espoo, Finland}
\affiliation{Institute for Theoretical Physics, Leibniz University of Hannover, Hannover, Germany}
\affiliation{Institute for Physics and Astronomy, TU Berlin, Germany}

\author{Maciej Lewenstein}
\affiliation{ICFO - Institut de Ci\`encies Fot\`oniques, The Barcelona Institute of Science and Technology, 08860 Castelldefels (Barcelona), Spain}
\affiliation{ICREA, Pg. Lluis Companys 23, ES-08010 Barcelona, Spain}

\author{Manabendra Nath Bera}
\email{mnbera@gmail.com}
\affiliation{Department of Physical Sciences, Indian Institute of Science Education and Research (IISER), Mohali, Punjab 140306, India}

\title{Fundamental Limitations on the Reliabilities of Power and Work in Quantum Batteries}

\begin{abstract}
Quantum batteries, microscopic devices designed to address energy demands in quantum technologies, promise high power during charging and discharging processes. However, their performance depends crucially on reliability, quantified by the inverse of noise-to-signal ratios (NSRs), i.e., normalized fluctuations of work and power. We establish fundamental limits to this reliability: both work and power NSRs are universally bounded from below by a function of charging speed, imposing a reliability limit inherent to any quantum battery. More strikingly, we find that a quantum mechanical uncertainty relation forbids the simultaneous suppression of work and power fluctuations, revealing a fundamental trade-off that also limits the reliability of quantum batteries. We analyze the trade-off and limits, as well as their scaling behavior, across parallel, collective, and hybrid charging schemes for many-body quantum batteries, finding that increasing power by exploiting stronger entanglement comes at the cost of diminished reliability of power. Similar trends are also observed in the charging of quantum batteries utilizing  Ising-like interactions. These suggest that achieving both high power and reliability requires neither parallel nor collective charging, but a hybrid charging scheme with an intermediate range of interactions. Therefore, our analysis shapes the practical and efficient design of reliable and high-performance quantum batteries.
\end{abstract}

\maketitle

\section{Introduction}
Rising global energy demands and technological advancement underscore the need for high-performance, reliable energy storage systems to power next-generation innovations. Quantum batteries (QBs), quantum-mechanical systems designed for work storage, possibly composed of quanta-cells, integrate principles of quantum physics and energy science to provide novel paradigms for powering quantum technologies~\cite{Alicki2013, Binder2018, Alexia2022, Quach2023, Kurman2025, Ferraro2026}. By exploiting superposition and entanglement, QBs can achieve superior charging power, efficiency, and scalability~\cite{Alicki2013, Campaiolo2017, Ferraro2018, Sergi2020, Mayo2022, Gyhm2022, Sugimoto2025}. Recent research has demonstrated their remarkable power output, ultra fast charging, and high work capacity~\cite{Binder2015, Mohan2022, Sergi2020, Mazzoncini2013, Gyhm2024, Mohan2025A}, along with enhanced stability and robustness~\cite{Rosa2020, Quach2020, Mohan2021, Grazi2025, Lu2025}. Theoretical studies continue to refine these systems by optimizing key performance metrics—power, efficiency, and scaling behavior—to facilitate practical realization~\cite{Sergi2020, Rossini2020, Konar2022, Shaghaghi2022, Shukla2025, Malavazi2026, Ahmadi2026}. Moreover, recent experimental advances have successfully demonstrated QBs across diverse physical platforms~\cite{Joshi2022, Quach2022,  Huang2023, Camposeo2025, Tibben2025, Hu2026}. For an in-depth overview of the field, see Ref.~\cite{Campaioli2024}.

Quantum batteries harness quantum dynamical properties to efficiently store and transfer energy during charging and discharging processes. However, these same processes inherently introduce unavoidable quantum dynamical fluctuations in both work and power. In fact, these fluctuations critically influence the battery’s performance, particularly the reliability of its work and power output. In general, the reliability of power (or work) is characterized by the noise-to-signal ratio (NSR), defined as the fluctuation in power (or work) normalized by its mean, where high NSR implies low reliability. 

It is worth noting that while a few studies have examined certain aspects of work fluctuations during battery-charging processes~\cite{Friis2018, McKay2018, Llobet2019, Crescente2020, Caravelli2020, Pintos2020, Bakhshinezhad2024, Campaioli2024, Imai2023, Sarkar2025, Rinaldi2025, Polo2025}, to the best of our knowledge, no prior work has investigated fluctuations in power. Recent studies have majorly focused on enhancing charging power, and several works have shown that entanglement generated in many-body batteries can potentially boost charging power \cite{Binder2015, Campaiolo2017, Sergi2020, Gyhm2022}. However, high power alone does not guarantee a high-performance quantum battery. Designing reliable quantum batteries by optimizing their performance need a clear understanding of the NSRs of work and power in relation to the charging processes. This, particularly, requires to address the following key questions: (i) how entanglement-assisted high-power charging influences the reliability of power and work; (ii) whether, for arbitrary unitary charging or discharging process, fundamental limits exist on the reliability of power and work; (iii) whether fundamental trade-offs arise between the reliabilities of power and work when one seeks to enhance them simultaneously; and (iv) how, in many-body quantum batteries, the reliabilities of power and work scale with the number of quanta-cells.

\begin{figure}
\centering

\includegraphics[width=8.4cm,height=5cm]{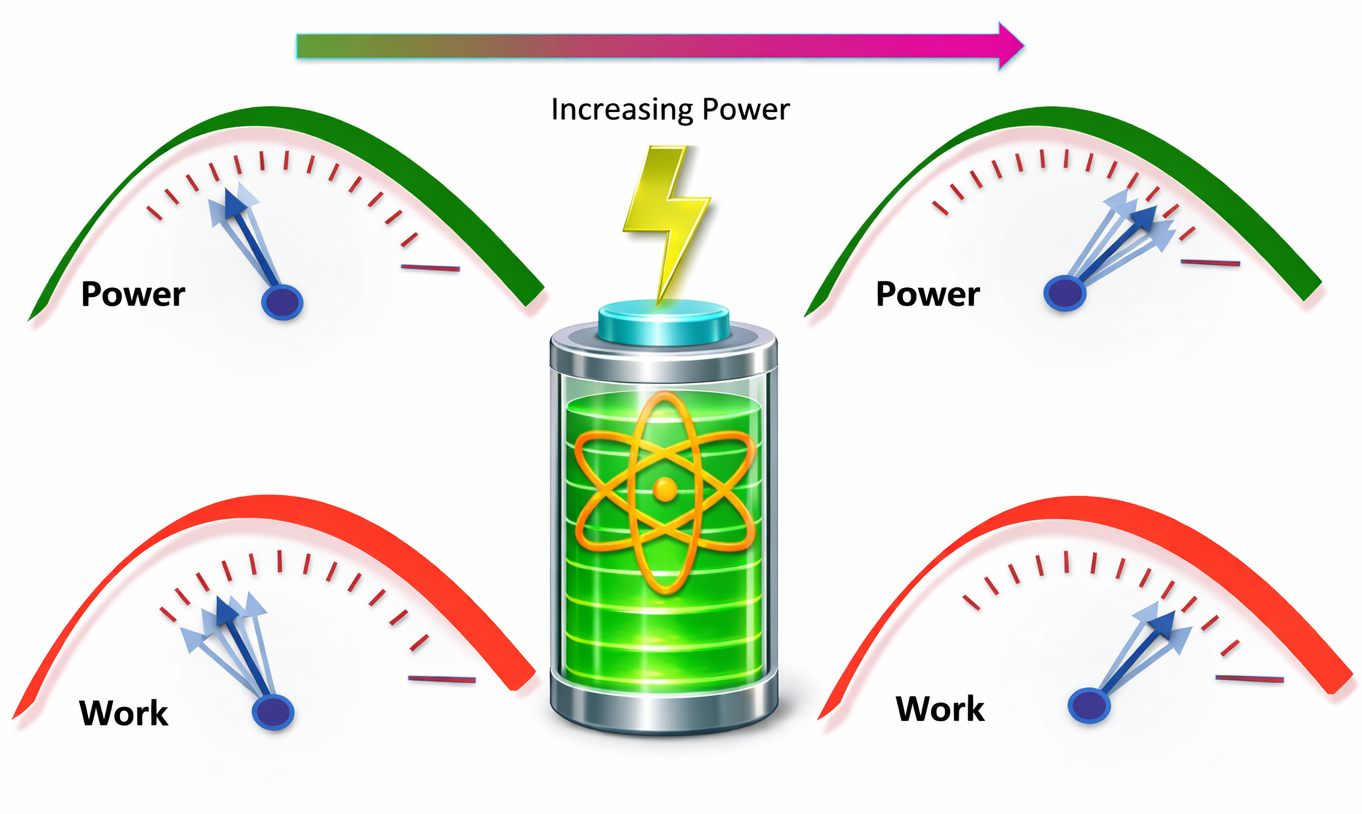}
\caption{\justifying{ \textbf{ Power and work vs their reliability in the charging process of quantum batteries.}
Schematic illustration of the reliability of a quantum battery during the charging process. The upward arrow denotes increasing charging power. At high power (right), the delivered power exhibits large fluctuations, indicated by multiple arrows in the power meter, which reduces the reliability of power. At the same time, fluctuations in the delivered work are suppressed, leading to a more stable reading in the work meter on the left. Conversely, in the low-power regime (left), the power meter shows smaller fluctuations, while the work meter exhibits larger variability. This opposite behavior highlights the intrinsic trade-off between the reliability of power and work during quantum battery charging.
 }}
\label{fig:diag}
\end{figure}

In this article, we attempt to address these questions systematically. We begin by establishing fundamental lower bounds on the NSRs of work and power, considered separately, for unitary charging and discharging processes. These bounds reveal that each NSR is inherently limited by the charging speed and a relevant time-correlation function. The NSRs of work and power can each be saturated to their lower bounds under specific charging protocols, both separately and simultaneously. However, the inherent quantum dynamical fluctuations, together with the non-commutativity of work and power operators,  give rise to a fundamental trade-off relationship between the NSRs of work and power, which is a direct consequence of the well-known Schr\"odinger–Robertson uncertainty relation. In turn, we show that work and power reliabilities are inherently incompatible during charging (or discharging) processes and cannot simultaneously be maximized to an arbitrary high value, as schematically illustrated in Fig.~\ref{fig:diag}. We further analyze how the NSRs scale with the number of quanta-cells in many-body quantum batteries under different charging strategies, including collective, parallel, and hybrid protocols, and we examine conditions under which the trade-off relation becomes saturated. Contrary to common expectation, our results demonstrate that high-power charging---often enabled by entanglement generation in many-body settings---typically leads to a significant reduction in reliability of power. Finally, we discuss the implications of these findings for a practical many-body quantum battery model and outline potential applications in designing and optimizing high-performance quantum batteries as well as promising directions for future research.\\

The article is organized as follows. In Section~\ref{sec:NSRs}, we introduce the normalized fluctuations (NSRs) of work and power for an arbitrary quantum battery undergoing unitary charging and discharging processes, using the full-counting statistics. Section~\ref{sec:LimitsAndTradeOff} presents the derivation and analysis of the main results, including: (i) fundamental lower bounds on the normalized fluctuations (NSRs) of power and work and the conditions under which these bounds are saturated; (ii) a fundamental trade-off relation between NSRs of power and work, derived from a modified quantum uncertainty relation, and their implications. In Section~\ref{sec:SclaingOfNSRs}, we study many-body quantum batteries and the behavior and scaling of NSRs, with particular emphasis on parallel, collective, and hybrid charging and discharging schemes, as well as spin-chain–based quantum batteries employing Ising-like interactions. Finally, we summarize our results and conclude in Section~\ref{sec:Conclusions}.

\section{Results}

\subsection{Quantum batteries and their reliability \label{sec:NSRs}}
A closed (noiseless) quantum battery, consisting of several non-interacting quantum mechanical systems described by the internal Hamiltonian $H^{B}_{0}$. It can store work and be charged by the external driving with the Hamiltonian $H_{t}^{C}$. Thus, the total Hamiltonian becomes
\begin{equation}
    H^{T} = H^{B}_{0} + H^{C}_t.
\end{equation}
Note, the external driving is switched on only during the charging (or discharging) process, from $t=0$ to $t=t_f$, i.e., $H^{C}_{t=0}=H^{C}_{t=t_f}=0$. In quantum batteries, the maximum extractable energy through a unitary process is defined as work (or ergotropy)~\cite{Allahverdyan2004}. The rate at which work is injected into or extracted from the battery is referred to as its charging and discharging power, respectively. However, both work and power can exhibit (quantum) fluctuations, which reflect the uncertainty in the amount of extractable work and the rate at which it can be delivered during the charging and discharging processes. 

The average and fluctuations of work and power during a unitary charging process can be evaluated at any time $t$ using full counting statistics (FCS), which aligns with two-point measurement (TPM) statistics formalism if initial state commutes with counting observable~\cite{Esposito2009}. Using the FCS formalism, the average and fluctuation (measured with variance) associated with any counting observable $O_0$ at time $t$ are given by
\begin{align}
    &\langle \cal{O}_t \rangle  = \Tr((O_{t}-O_{0})\rho_0),\label{DefW/P} \\ \ \mbox{and} \ \nonumber \\
  & ( \Delta  \cal{O}_t)^2   =\Tr((O_{t}-O_{0})^2\rho_0) -\langle \cal{O}_t \rangle^2,\label{DefP}
\end{align}
respectively, where $\rho_0$ is the initial state of battery, the observable \(O_t = \exp(i H^T t / \hbar) O_0 \exp(-i H^T t / \hbar)\) and $ \cal{O}_t  =O_{t}-O_{0}$. See Supplementary Material~\ref{FCS} for more details. Now, the reliability of the quantities corresponding to (counting) observable $O_0$, during the charging or discharging process, is given by its noise-to-signal ratio (NSR), i.e., 
\begin{equation}\label{nsr}
 \cal{N}_t^{\cal{O}}: = \frac{(\Delta \cal{O}_t)^2 }{\langle \cal{O}_t  \rangle^2}.
\end{equation}

 It is important to note that the average value, fluctuations, and noise-to-signal ratio (NSR) of the observable \(\mathcal{O}_t\) are defined over a finite time interval \([0, t]\) and are therefore inherently two-time (or two-point) quantities.
By considering energy observable ( i.e., internal Hamiltonian of battery) $O_0=W_0=H^B_0$ as the counting observable for work, we compute the average, fluctuation, and NSR of work as $\langle \cal{W}_t \rangle$ and $\Delta \cal{W}_t$, and $\cal{N}_t^{\cal{W}}$, respectively. Since the rate of energy change of the battery is estimated using the expectation value of 
$-\frac{i}{\hbar}[H^B_0, H^C_t]$ in the time-evolved battery state $\rho_t=\exp(-i H^T t / \hbar) \rho_0 \exp(i H^T t / \hbar)$, we can define a counting observable $O_0 = P_0 = -\frac{i}{\hbar}[H^B_0, H^C_t]$ to compute the average, fluctuations, and noise-to-signal ratio (NSR) of power, denoted respectively as 
$\langle \mathcal{P}_t \rangle$, $\Delta \mathcal{P}_t$, and $\mathcal{N}_t^{\mathcal{P}}$. The counting observables $W_0$ and $P_0$,  use to count averages and fluctuations in work and power, are generally non-commuting for any nontrivial charging Hamiltonian. This feature has significant consequences for the stability and reliability of quantum batteries as we discuss below.\\

 It is worthwhile to mention that, in the closed quantum-battery setting, both work and power can be represented by Hermitian operators, as given in Eq.~\eqref{DefW/P}. This is because the battery evolves unitarily under a prescribed charging Hamiltonian, and there is no heat exchange with an environment. In this case, the work performed on the battery is identified with the change in its internal energy, while the corresponding power is obtained from the rate of change of this internal-energy difference. The averages and variances of work and power obtained from this operator-based definition (given in Eqs.~~\eqref{DefW/P} and ~\eqref{DefP}) are consistent with those obtained from the two-point measurement scheme~\cite{Allahverdyan2005, Esposito2009, Allahverdyan2014, Solinas2015, Talkner2016}, when the initial state commutes with the corresponding counting observable, and with the full-counting-statistics formalism~\cite{Esposito2009, Solinas2015, Solinas2016,  Xu2018, Francica2022, Francica2023, Santiago2024}. Therefore, the averages and variances obtained in this section are not specific to any particular formalism used to define them. Accordingly, work, power, and their respective fluctuations are defined using widely accepted techniques in quantum thermodynamics~\cite{Allahverdyan2005, Esposito2009, Allahverdyan2014, Solinas2015, Talkner2016, Solinas2016,  Xu2018, Francica2022, Francica2023, Santiago2024}. Moreover, the definition of work is also consistent with the notion of ergotropy widely used in the quantum-battery literature~\cite{Allahverdyan2004, Campaioli2024}.

However, this operator-based description of work and power is specific to closed quantum systems, in which the change in the system’s energy is governed by a unitary process, as considered in this work. In general, especially for open quantum systems interacting with an environment, work and the associated power are path-dependent quantities. Therefore, they cannot, in general, be associated with Hermitian operators~\cite{Talkner2007, Allahverdyan2014, Hovhannisyan2024, Silva2024, Rubino2025}.

It is important to distinguish the two-time NSRs of work and power given by Eq.~\eqref{nsr} from the instantaneous relative fluctuations of the battery energy and its rate of change during the charging or discharging process. The two-time quantities characterize energy transfer over a finite time interval \([0, t]\), whereas the instantaneous quantities define at a particular instant of time $t$. At a given time $t$, the instantaneous relative fluctuation of an observable \({O}_0\) is defined as
\( 
\frac{\left(\Delta {O}_t\right)^2}{\langle {O}_t\rangle^2},
\)
where \(\langle {O}_t\rangle\) and \(\left(\Delta {O}_t\right)^2\) denote its mean and variance in state $\rho_0$, respectively at any time $t$. For \({O}_0={W}_0=H_0^B\), the corresponding time-evolved counting observable \({W}_t\) represents the time evolved battery Hamiltonian at time $t$. Therefore,
\( 
\frac{\left(\Delta {W}_t\right)^2}{\langle {W}_t\rangle^2}
\)
characterizes the instantaneous relative fluctuation in the battery energy. Similarly, for \({O}_0={P}_0\), the corresponding time-evolved observable \({P}_t\) represents the instantaneous energy rate operator, and
\( 
\frac{\left(\Delta {P}_t\right)^2}{\langle {P}_t\rangle^2}
\)
characterizes the instantaneous relative fluctuation in the rate of change of battery energy. Since, work described as the change in the battery energy over the finite interval \([0,t]\), whereas power characterizes the corresponding rate of energy transfer over that interval. Accordingly, the two-time NSR of work quantifies the relative fluctuations in the energy deposited in or extracted from the battery during \([0,t]\), whereas the two-time NSR of power quantifies the relative fluctuations in the corresponding energy-transfer rate. The two-time NSRs depend on the fluctuations at both endpoints, $t_i=0$ and $t_f=t$, together with the correlations between them, on the another end instantaneous relative fluctuation depends only on instantaneous time $t_f=t$. Thus, these instantaneous relative fluctuations are distinct from the two-time NSRs of work and power given by Eq.~\eqref{nsr}. Moreover, two-time NSRs retain more dynamical information about the charging or discharging process that is absent from purely instantaneous relative fluctuations. In the context of quantum batteries, the relevant quantities are work and power, rather than the average energy of the battery and its corresponding rate of change. Therefore, in this work, we consider the two-time NSRs of work and power defined in Eq.~\eqref{nsr}, since work and power are naturally defined as two-time quantities.

The NSRs of work and power provide complementary measures of reliability while the quantum battery is actively operating. The NSR of work quantifies the precision with which energy is stored in or extracted from the battery, whereas the NSR of power quantifies the stability of the corresponding energy-transfer rate. Therefore, for a battery to be operationally useful, the fluctuations in both work and power should remain small relative to their respective mean values, thereby ensuring a reliable energetic output and a stable energy-transfer rate throughout the charging or discharging process.

Once the charging or discharging process has been switched off, the uncertainty in the final stored or extracted energy characterizes the endpoint reliability of the battery. However, in this work, we focus on the reliability of the quantum batteries while it is actively operating, namely, during the charging or discharging process, as quantified by the two-time NSRs of work and power defined in Eq.~\eqref{nsr}.

\subsection{Fundamental limitations on reliabilities \label{sec:LimitsAndTradeOff}}
In general, low NSRs of both work and power during charging and discharging are essential for the stable and reliable operation of quantum batteries. Therefore, designing and optimizing the performance of quantum batteries requires understanding the fundamental limits on the NSRs of work and power, separately, as well as identifying any potential trade-offs when attempting to minimize NSRs of both simultaneously.

\subsubsection{Fundamental lower bounds on NSRs}
Ideally, highly reliable quantum batteries are expected to exhibit vanishing NSRs at all times, indicating perfect reliability, for work and power separately. However, due to fundamental dynamical constraints of quantum mechanics, the NSRs of work and power cannot be reduced to zero. We obtain the fundamental lower bounds on the reliability of work and power at time $t$, given as ($\cal{O}=\{\cal{W}, \cal{P}\}$)
\begin{equation}\label{NSRBB}
\cal{N}_t^{\cal{O}} \geq {\cot^2\!\left(\int_{0}^{t}dt^{'} {{{\sqrt{\cal{I}^{O_0}_{t'}}}{\Big/}{2}}} \!\right)}- f(O_0,O_t).
\end{equation}
The classical Fisher information $\cal{I}^{O_0}_{t} \geq 0$ quantifies the speed of charging when the battery state is projected onto the eigenspace of $O_{0}$~\cite{Sergi2020}. The correlation function $f(O_0, O_t) \geq 0$ represents the correlation function between $O_0$ and $O_{t}$ and becomes zero if the initial battery state is an eigenstate of the counting observable $O_0$. For the derivation of Eq.~\eqref{NSRBB} and other details, see Supplementary Material~\ref{PLB}. The bounds in Eq.~\eqref{NSRBB} suggest that NSR of work (and power) is constrained by a function of the charging speed and the correlation function. This implies that the NSRs of both work and power cannot be reduced to arbitrarily low values for any unitary charging or discharging process; in other words, arbitrarily high reliability cannot be achieved. The lower bounds in NSRs are time-dependent and may vary during charging process. Nevertheless, for certain processes with some initial states and charging Hamiltonians, the NSRs of work and power can be simultaneously minimized down to their fundamental lower bounds. One such case is charging of single qubit quantum battery presented in Supplementary Material~\ref{SQB}.

\subsubsection{Fundamental trade-off between Reliabilities}
A reliable quantum battery must aim to minimize the NSRs of both power and work at every instant of time. Different charging schemes, characterized by distinct charging Hamiltonians and initial battery states, can be chosen such that the NSRs of either power or work become arbitrarily small. However, due to the intrinsic incompatibility of quantum observables, it is generally not possible to achieve simultaneously low NSRs of both work and power during the charging and discharging processes of a quantum battery, as we now discuss.

Note that the power measures the rate of energy (work) transfer into or out of a quantum battery, and it is related to $-\frac{i}{\hbar}[H_0^B, H_t^C]$. Therefore, while quantifying the average value or fluctuation in power, the counting observable associated with power is $-\frac{i}{\hbar}[H_0^B, H_t^C]$, and the counting observable corresponding to work, $H_0^B$, are generally incompatible. Therefore, for this reason, the NSRs of work and power satisfy a trade-off relation given by
\begin{equation}\label{TOWP}
  \mathcal{N}_t^\mathcal{W} \mathcal{N}_t^\mathcal{P}\geq 
 \frac{1}{4} (| \langle[\tilde{\mathcal{P}_t}, \tilde{\mathcal{W}_t}]\rangle|^{2}
  +  | \langle\{\tilde{\mathcal{P}_t}, \tilde{\mathcal{W}_t}\}\rangle - 2\rangle|^{2}),
\end{equation}
where $\mathcal{N}_t^\mathcal{W}$ and $\mathcal{N}_t^\mathcal{P}$ denote the NSRs of work and power, respectively, and $\tilde{\mathcal{P}_t} = \frac{\mathcal{P}_t}{\langle \mathcal{P}_t \rangle}$ and $\tilde{\mathcal{W}_t} = \frac{\mathcal{W}_t}{\langle \mathcal{W}_t \rangle}$ can be regarded as the normalized Heisenberg operators of power and work~\cite{Allahverdyan2005}. The $[\cdot, \cdot]$ and $\{\cdot, \cdot\}$ represent the commutation and anti-commutation respectively and \(
\langle X \rangle = \Tr(X \rho_0)\). Note that the trade-off relation above can be interpreted as a variant of the well-known Schrödinger–Robertson uncertainty relation~\cite{Robertson1929, Schrodinger1930}. Moreover, in a similar way, another trade-off relation may be expressed as a sum of NSRs using the $AM$-$GM$ inequality~\cite{Gaines1967} or sum uncertainty relation~\cite{Maccone2014}, i.e., $\cal{N}_t^\mathcal{W} + \cal{N}_t^\mathcal{P} \geq 2 \sqrt{\cal{N}_t^\mathcal{W} \cal{N}_t^\mathcal{P}}$. The trade-off relation in Eq.~\eqref{TOWP} provides a universal constraint or incompatibility of the reliability in work and power in all closed quantum batteries. Thus, attempting to reduce the NSR of power inevitably comes at the cost of an increased NSR of work, or vice versa, at least when the trade-off relation is saturated.  For the derivation of Eq.~\eqref{TOWP} and other details, see Supplementary Material~\ref{FTFR}. It is important to note that, by dropping the terms involving \({O}_0\) from the bounds in Eqs.~\eqref{NSRBB} and~\eqref{TOWP}, we obtain the corresponding instantaneous bounds on the relative fluctuations of the battery energy and its rate of change during the charging and discharging processes. In this scenario, Eq.~\eqref{TOWP} yields a trade-off relation between the relative fluctuations of the battery energy and the relative fluctuations of its instantaneous rate of change, both evaluated at time $t$. Such a relation reflects the incompatibility between the instantaneous fluctuations of the battery energy and its rate of change; see, for example, Ref.~\cite{Sathe2024}. Importantly, our trade-off captures the fluctuations of the entire charging process rather than just the instantaneous fluctuations. Hence, it is more general.

In recent times, several studies have focused on enhancing the power output of quantum batteries~\cite{Binder2015, Campaiolo2017, Sergi2020, Gyhm2022, Campaioli2024} by leveraging many-body quantum effects, such as  entanglements , which can accelerate quantum evolution and thereby increase battery power. However, in general that enhancing power may lead to high fluctuations in it. Nevertheless, power and its reliability alone are not the sole metric of interest; the work and its reliability are also equally important. Our trade-off relation reveals that enhancing the reliability of power typically comes at the cost of decreased reliability in the delivered work, and vice versa. In fact, an efficient quantum battery must achieve high reliability in both power and work, in addition to maintaining high power output.

\subsection{Many-body quantum batteries \label{sec:SclaingOfNSRs}}
The bounds on the NSRs derived in Eq.~\eqref{NSRBB}, together with the trade-off relation in Eq.~\eqref{TOWP}, are fundamental and model-independent, and therefore apply to arbitrary closed quantum battery architectures. We now employ these results to investigate the performance of many-body quantum batteries using the NSRs and the other figures of merit introduced above.

For this, we consider two classes of models. The first consists of paradigmatic models that allow us to clearly elucidate how quantum entanglement generated during the charging and discharging processes affects work, power, fluctuations, NSRs, and the work–power reliability trade-off. In particular, we examine the scaling behavior of NSRs with the number of quanta-cells and assess whether quantum entanglement truly enhances battery reliability. The second class comprises experimentally accessible battery models that utilize Ising-like interactions for charging and discharging processes, which, informed by insights gained from the paradigmatic models, may guide the design of high-performance quantum batteries.

\subsubsection{Paradigmatic models}
We start with paradigmatic models of (many-body) quantum batteries consisting of $N$ non-interacting spin-$\frac{1}{2}$ particles. Each of these spin particles can be regarded as a quanta-cell constituting the battery. The Hamiltonian of the battery is given by $H_{0}^{B} = -\frac{\omega_0}{2} \sum_{i=1}^{N} \sigma_z^{(i)}$, where $\sigma_z$ denotes the Pauli-$z$ matrix and $\omega_0$ is the energy splitting. Here, we set $\hbar=1$, and denote $\ket{0}$ and $\ket{1}$ as the ground and excited state of each quanta-cell, respectively. The charging (or discharging) of the battery is driven by an external field, described by the Hamiltonian $H^C_{t,k}$. Thus, the total Hamiltonian during the charging (or discharging) process becomes $H^T=H_{0}^{B} + H^{C}_{t,k}$. Here, the charging Hamiltonian $H^{C}_{t,k}$ is not necessarily local and may introduce $k$-spin interaction between the quanta-cells during the charging or discharging process. Without loss of generality, we assume $\frac{N}{k} \in \mathbb{N}$. Then, a generic expression of this charging Hamiltonian is 
\begin{equation}
H_{k,t}^{C} =  {\Omega_{k}} \sum_{j=0}^{\frac{N}{k}-1} \bigotimes_{i=1}^k \sigma_x^{(k j + i)}
 - H_{0}^{B}. 
 \label{eq:HcManyBody}
\end{equation}
The cases with $\frac{N}{k} \notin \mathbb{N}$ are considered in Supplementary Material~\ref{sec:k body hamiltonian}. Nevertheless, depending on the value of $k$, this Hamiltonian represents different levels of interaction between the spins. For example, when $k = 1$, the Hamiltonian $H_{1}^{C} = \Omega_{1} \sum_{i=1}^N \sigma_x^{(i)} - H_{0}^{B}$ describes a scenario in which each quanta-cell is charged (or discharged) independently. This is referred to as parallel (local) charging. At the other end, for $k = N$, the Hamiltonian $ H_{N}^{C} = \Omega_{N} \bigotimes_{i=1}^N \sigma_x^{(i)} - H_{0}^{B}$ represents a fully collective (or fully non-local) process where all quanta-cells interact with one another. This is denoted as {\it collective charging}. There are also intermediate cases for $1 < k < N$, where the charging process involves $k$-body interactions, as described by the Hamiltonian in Eq.~\eqref{eq:HcManyBody}. We denote this process as the hybrid (semi-local) charging process. For a  fair comparison between various charging schemes, we impose the condition $\norm{H_{k,t}^T}=\Omega_{0}, \ \forall \ k$, where $\Omega_{0}$ is constant. This further reduces to the condition, $\Omega_{k} =\frac{k}{N}\Omega_{0}, \ \forall \ k$. Note, $\Omega_k \leq \Omega_{k+1}$ for a fix $N$. \\

\begin{figure*}
\centering
\includegraphics[width=0.33\textwidth]{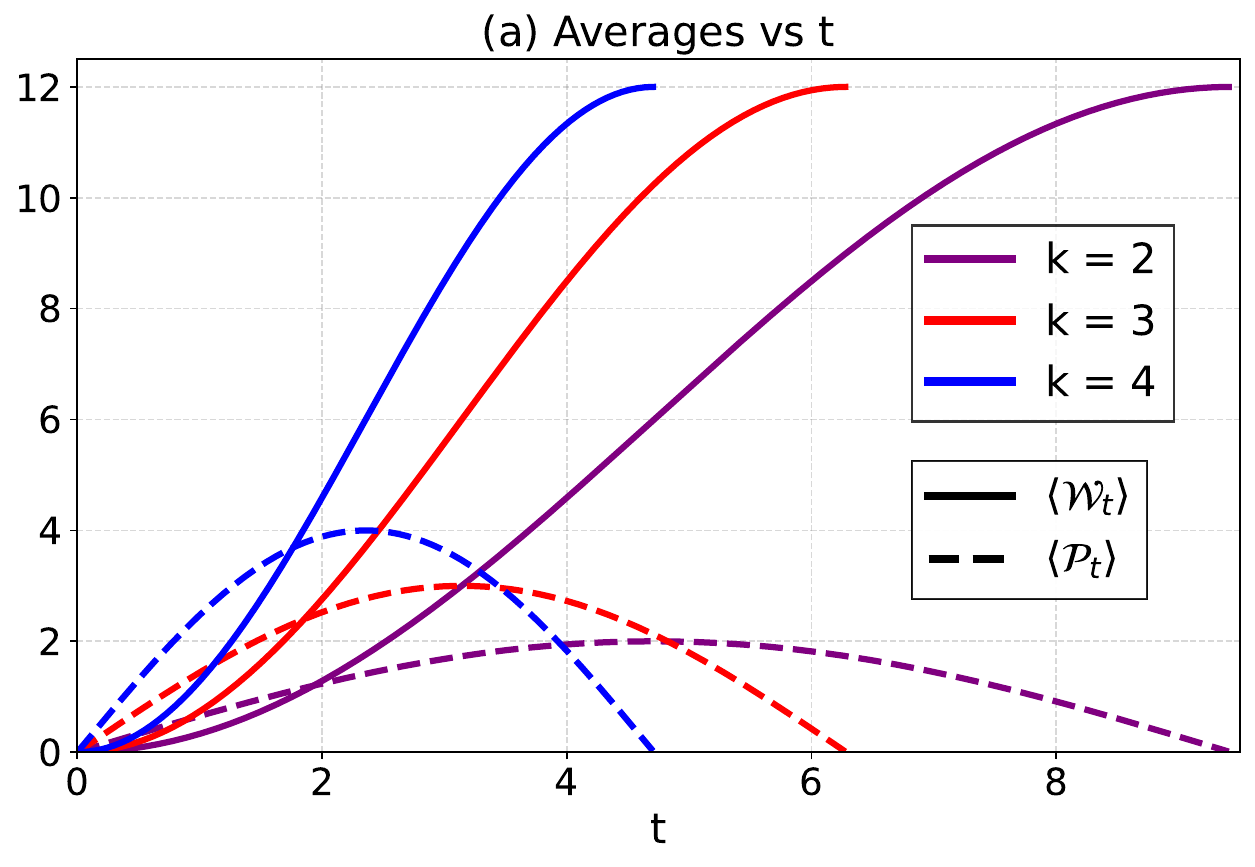}
\includegraphics[width=0.33\textwidth]{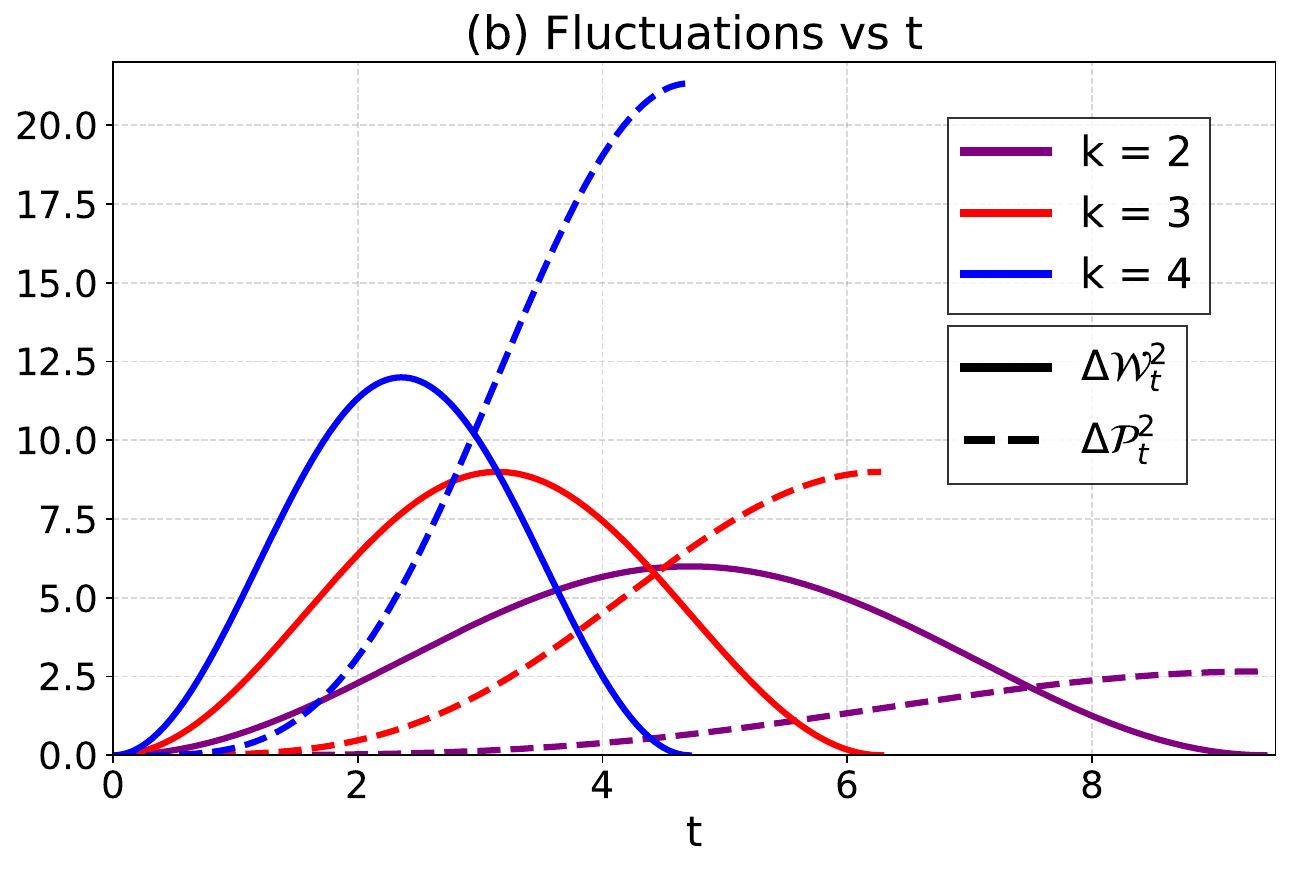}
\includegraphics[width=0.33\textwidth]{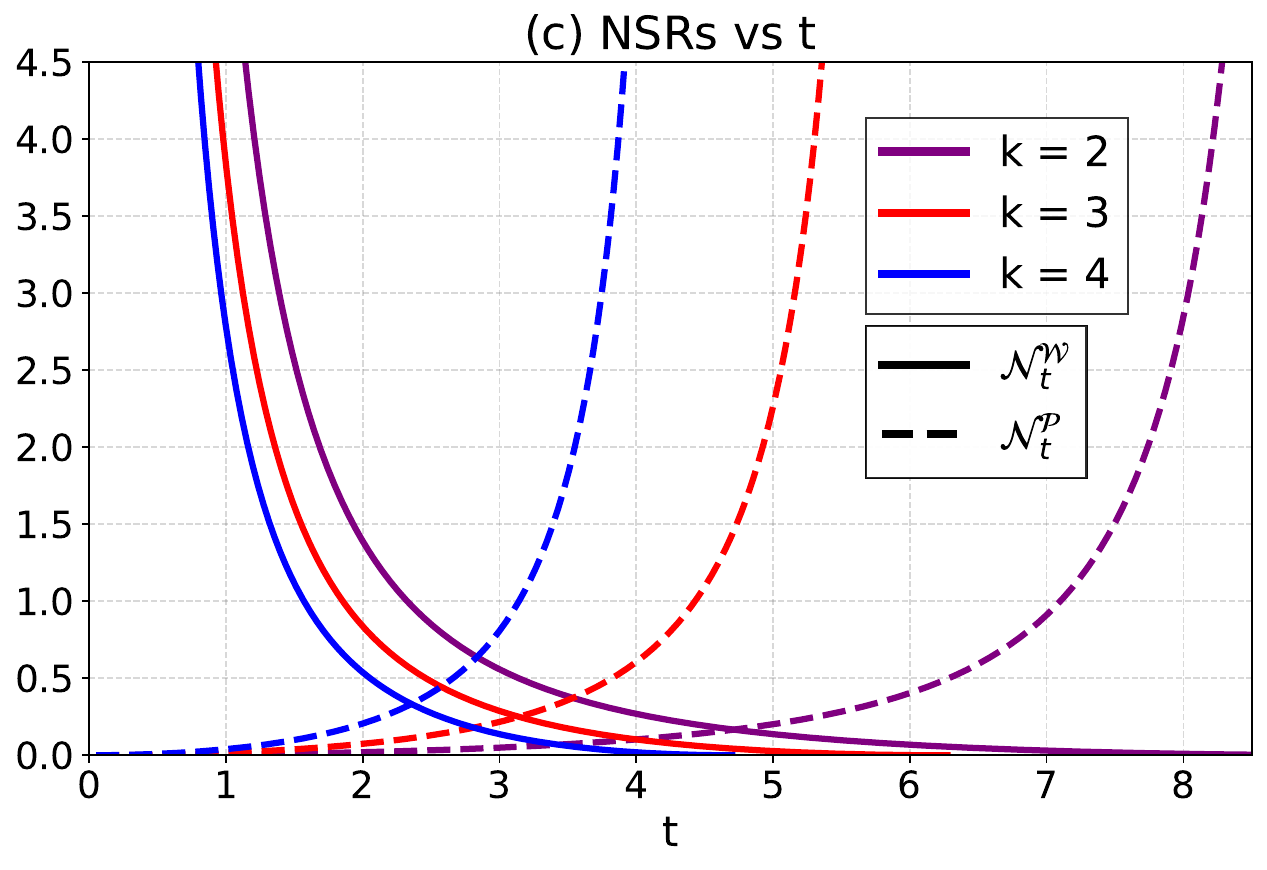}
\caption{\justifying{The plots (a)–(c) illustrate the charging dynamics of a quantum battery governed by a $k$-body charging Hamiltonian given in Eq.~\eqref{eq:HcManyBody}, with $N = 12$, $\Omega_0 = 1$, and $\Omega_k = \tfrac{k}{N}\Omega_0$. In all Plots, $k =2, \ 3, \mbox{and}$ 4 represented by purple, red, and blue curves, respectively. Plot (a) shows the average work $\langle \mathcal{W}_t^{(k)} \rangle = N\omega_0\sin^2(\Omega_k t)$ (solid lines) and average power $\langle \mathcal{P}_t^{(k)} \rangle = k\omega_0\Omega_0\sin(2\Omega_k t)$ (dashed lines). 
Plot (b) depicts the corresponding fluctuations $(\Delta \mathcal{W}_t^{(k)})^2 = Nk\omega_0^2\sin^2(2\Omega_k t)/4$ (solid) and $(\Delta \mathcal{P}_t^{(k)})^2 = 4\frac{k^3}{N}(\omega_0\Omega_0)^2\sin^4(\Omega_k t)$ (dashed). 
(c) depicts the NSRs of work and power, $\mathcal{N}_t^{\mathcal{W},(k)} = \tfrac{k}{N}\cot^2(\Omega_k t)$ (solid) and $\mathcal{N}_t^{\mathcal{P},(k)} = \tfrac{k}{N}\tan^2(\Omega_k t)$ (dashed), shows their trade-off relation. }}
\label{fig:kbodyplot}
\end{figure*}

For the initial (discharged) state of the battery $\ket{\psi_0}=\ket{0}^{\otimes N}$, the time-evolved state becomes 
\begin{align*}
\ket{\psi_t^{(k)}} = \bigotimes_{j=1}^{\frac{N}{k}} [\cos{(\Omega_k t)}|0\rangle^{\otimes k} -i \sin{(\Omega_k t)}|1\rangle^{\otimes k}]_j.    
\end{align*}
Clearly, while the time-evolved state $\ket{\psi_t^{(1)}}$, for $k=1$, corresponding to parallel charging, remains uncorrelated throughout the evolution, the time-evolved state $\ket{\psi_t^{N}}$, for $k=N$, becomes $N$-party (genuine) entangled in the case of collective charging. It is straightforward to see that, for $1 < k < N$, the time-evolved state $\ket{\psi_t^{k}}$ can exhibit $k$-party entanglement at most. The detailed calculations of the relevant quantities considered henceforth are provided in Supplementary Material~\ref{sec:k body hamiltonian}. At any time $t$, the instantaneous power and its fluctuations, for $\beta=\omega_0 \Omega_0$, are $\langle \cal{P}_t^{(k)} \rangle = k  \beta  \sin(2\Omega_k t)$ and $(\Delta \cal{P}^{(k)}_t)^2 = \frac{4k^3\beta^2 }{N}  \sin^4(\Omega_k t)$, respectively. Also, the instantaneous work and its fluctuations respectively are $\langle \cal{W}_t^{(k)} \rangle = N  \omega_0 \sin^2(\Omega_k t)$ and $(\Delta \cal{W}^{(k)}_t)^2 = \frac{ N k \omega_0^2 }{4} \sin^2(2\Omega_k t)$. From these, we compute the NSRs of work $\mathcal{N}_t^{\mathcal{W},(k)}$ and NSR of power $\mathcal{N}_t^{\mathcal{P},(k)} $, given as
\begin{equation}~\label{Scaling}
    \mathcal{N}_t^{\mathcal{W},(k)} = \frac{k}{N}\cot^2\!\left(\Omega_k t\right),
\ \mbox{and} \
\mathcal{N}_t^{\mathcal{P},(k)} = \frac{k}{N}\tan^2\!\left(\Omega_k t\right).
\end{equation}

Understanding battery efficiency and performance employing NSRs and their scaling has far-reaching implications, presenting a complementary yet contrasting perspective on efficient quantum batteries compared to previous studies. For instance, studies in Refs.~\cite{Binder2015,  Campaiolo2017, Sergi2020, Gyhm2022, Campaioli2024} have shown that a charging Hamiltonian capable of generating multipartite entanglement during the charging process can enhance the charging power, a phenomenon often referred to as the \emph{quantum advantage}, which is corroborated by our observation as well.  It is worth noting that an enhancement of charging power in a collective charging process is often associated with the dynamical generation of multipartite entanglement; however, the converse does not necessarily hold  \cite{Gyhm20240, Shukla2026}.

For a fixed $N$, as illustrated in Fig.~\ref{fig:kbodyplot}(a), when the charging scheme changes from the parallel to the collective regime, i.e., $\frac{k}{N} \to 1$, the period of charging $T=\frac{\pi N}{2k \Omega_0}$ decreases, and the emergence of (multi-partite) entanglement leads to a significant enhancement in power relative to the parallel case. Nevertheless, a crucial aspect that has not been examined so far is the behavior of power fluctuations. We observe that these fluctuations can, in fact, transform the apparent quantum advantage into a \emph{disadvantage}: as $\frac{k}{N} \to 1$, the power fluctuation $(\Delta \mathcal{P}^{(k)}_t)^2$ grows substantially faster than average power (see Figs.~\ref{fig:kbodyplot}(a) and \ref{fig:kbodyplot}(b)). As the dominant scaling (ignoring the trigonometric factors) shows that the fluctuation increases as $k^3$, whereas the average power scales only linearly with $k$. Consequently, the reliability of the power output of the quantum battery drops sharply, i.e., $\mathcal{N}_t^{\mathcal{P},(k)} = \tfrac{k}{N}\tan^2(\Omega_k t)$, increases as $\frac{k}{N} \to 1$ (see Fig.~\ref{fig:kbodyplot}(c)). This reveals that charging schemes generating stronger entanglement are associated with lower reliability of power. Viewing it differently, for a fixed value of $k$, increasing $N$ leads to a reduction in both the average and fluctuation of power. In the limit $N \to \infty$, with $\frac{k}{N} \to 0$, the charging protocol effectively approaches parallel charging, thereby enhancing reliability of power. This shows that the generation of strong entanglement (with high $k$) during the charging or discharging process significantly compromises reliability in power, although it may yield high power output.

The behavior of the work output, for a fixed $N$, displays a different behavior, though the average work depends weakly on $k$, through $\Omega_k$ in the trigonometric factor. The fluctuation in work increases as $\frac{k}{N} \to 1$, which grows linearly with ${k}$ (ignoring the trigonometric factors). Consequently, as shown in Fig.~\ref{fig:kbodyplot}, the reliability of the instantaneous work increases, i.e., $\mathcal{N}_t^{\mathcal{W},(k)} = \tfrac{k}{N}\cot^2(\Omega_k t)$ decreases as $\frac{k}{N} \to 1$. Hence, charging protocols that produce higher degrees of entanglement are likely advantageous in terms of work reliability. For fixed $k$, both the average work and its fluctuations linearly increase with $N$. Consequently, the NSR of work increases in the limit $N \to \infty$, i.e., for $\frac{k}{N} \to 0$ (parallel charging cases). Therefore, it is opposite to the behavior of reliability of power, a higher degree of entanglement (corresponding to large $k$) generated during the charging or discharging process significantly increases the reliability of work.

\begin{figure*}
\centering
\includegraphics[width=0.33\textwidth]{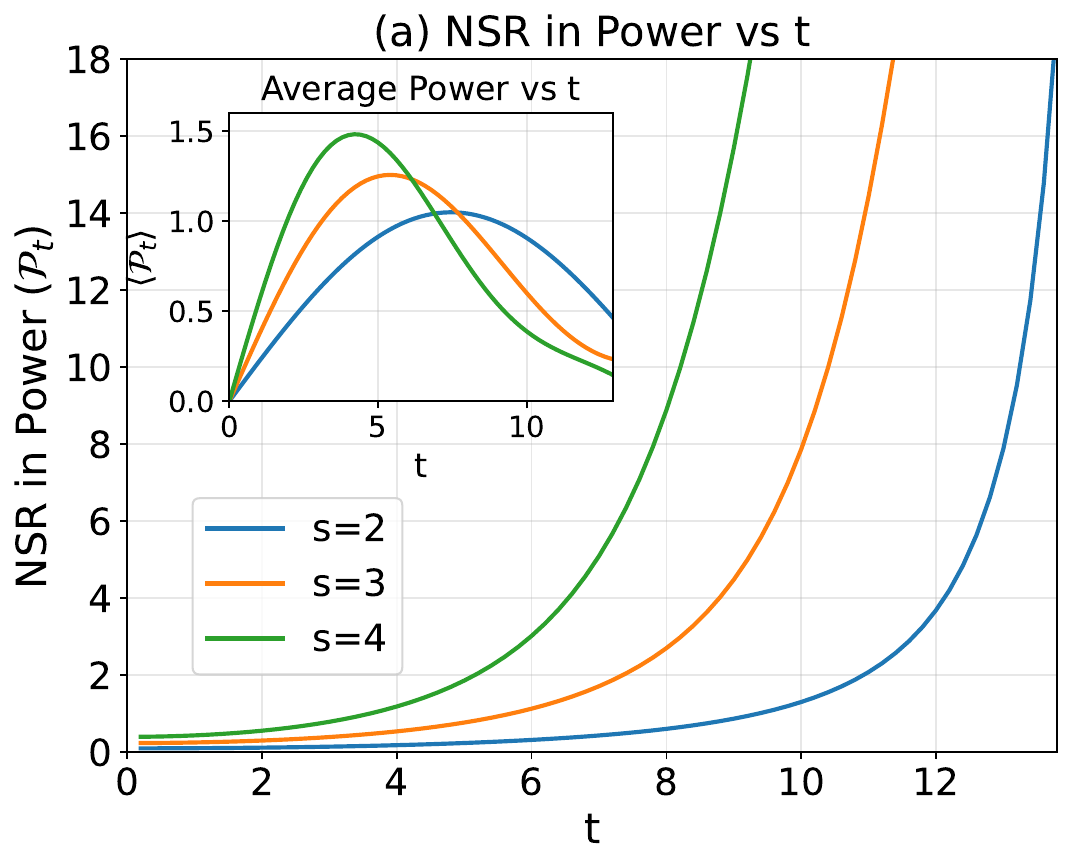}
\includegraphics[width=0.33\textwidth]{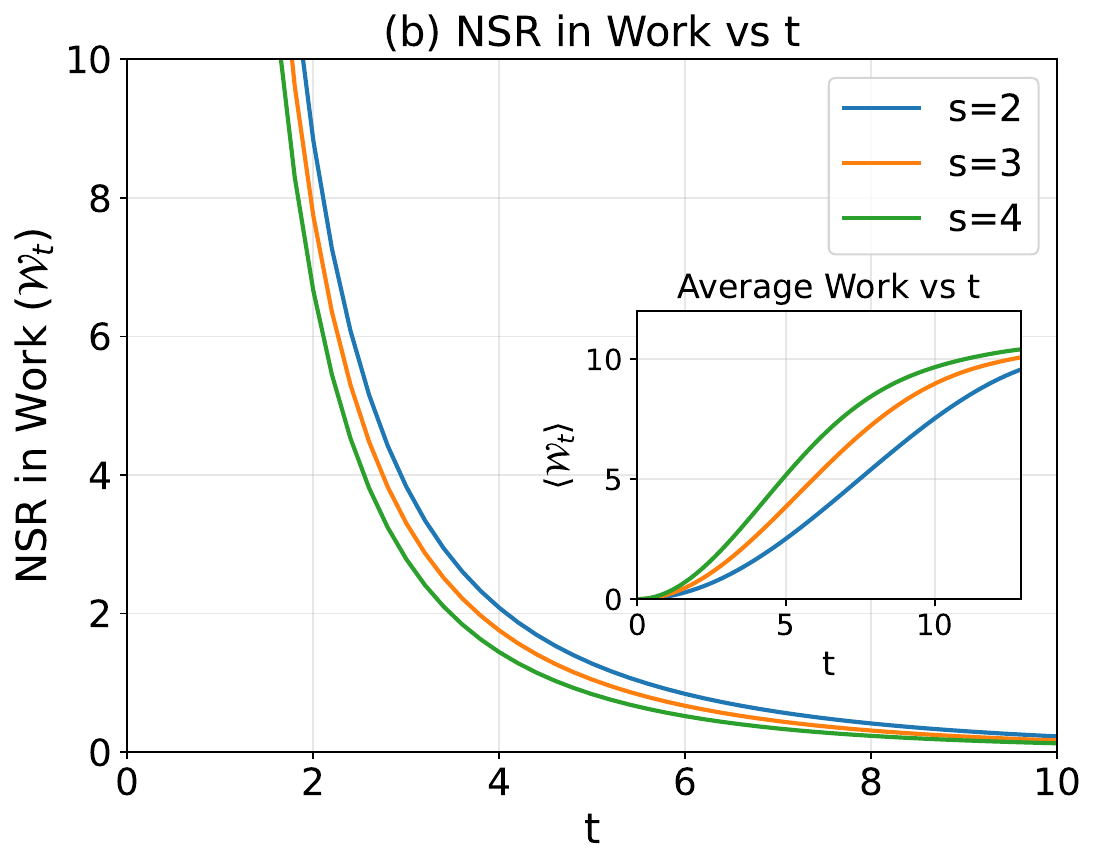}
\includegraphics[width=0.33\textwidth]{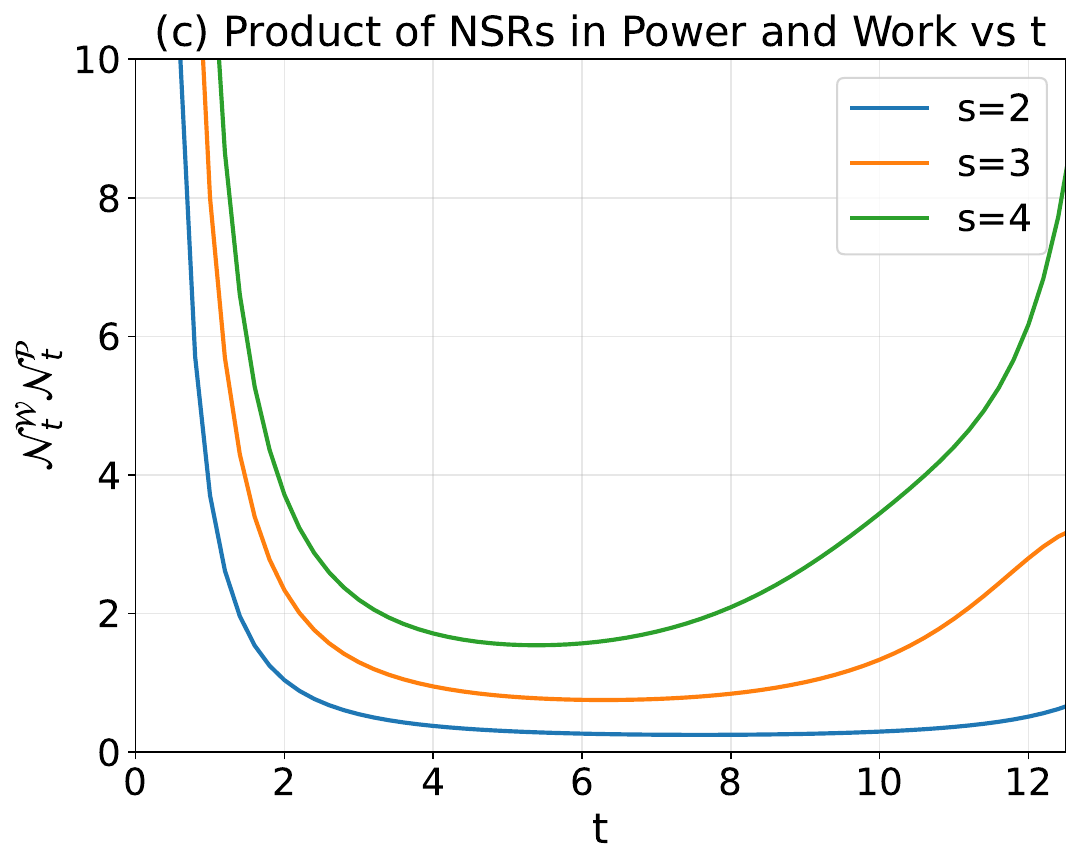}
\caption{\justifying{ Panels (a)–(c) illustrate the charging dynamics of a quantum battery governed by an $s$-body charging Hamiltonian given in Eq.~\eqref{XX}, for $N = 10$, $\Omega_s = 1$, and $\omega_0 = 2$. All plots are generated for $s = 2, 3,$ and $4$, corresponding to two-, three-, and four-body interactions, and are represented by blue, orange, and green curves, respectively. Panel (a) shows the power and the NSR of power, panel (b) presents the average work and the NSR of work, and panel (c) displays the product of the NSRs of work and power, $\mathcal{N}_t^{\mathcal{W},(s)} \mathcal{N}_t^{\mathcal{P},(s)}$. The plots of work and power fluctuations are provided in Supplementary Material~\ref{transverse}.
}}
\label{XX-Plots}
\end{figure*}

Thus on the basis of the above analysis, when considering the reliabilities in  work and power, the scaling relation~\eqref{Scaling} reveals that, contrary to common intuition, collective charging is actually less advantageous than parallel charging for achieving high reliability in power, on the other hand it is advantageous for achieving high reliability in work. This opposite behavior of reliability in work and power can also been explicitly observe if we consider the case when $N/k>>\Omega_0 t$, then quick calculation reveals that $\mathcal{N}_t^{\mathcal{P},(k)} \propto (\frac{k}{N})^3$ and $\mathcal{N}_t^{\mathcal{W},(k)} \propto \tfrac{N}{k}$ (see Supplementary Material~\ref{sec:k body hamiltonian}). Note, while the expression NSRs, $\mathcal{N}^{\mathcal{W},(k)}_{t}$ and $\mathcal{N}^{\mathcal{P},(k)}_{t}$ (see Eq.~~\eqref{Scaling}), are time-dependent, their product becomes independent of time. Most importantly, the product of NSRs of work and power exhibits a \emph{universal cluster scaling}:
\begin{equation}\label{eq:NSRScaling}
\mathcal{N}^{\mathcal{W},(k)}_{t}\mathcal{N}^{\mathcal{P},(k)}_{t}  = \frac{k^2}{N^2}.
\end{equation}  
Moreover, the trade-off relation Eq.~~\eqref{TOWP} is saturated for arbitrary value of $k$ and $N$, which suggest that reliabilities in work and power exhibits a genuine trade-off relation here (see Supplementary Material~\ref{sec:k body hamiltonian}). Thus, as $\frac{k}{N} \to 1$, i.e., towards fully collective charging, the product of NSRs becomes independent of $N$. In contrast, for $\frac{k}{N} \to 0$, i.e., towards parallel charging, it approaches zero as $N \to \infty$.

Therefore, parallel charging provides the highest possible reliability in power for many-body quantum batteries, albeit at the expense of significantly reduced power and reliability in work or vice-versa.  
These observations indicate that a truly efficient quantum battery—one that seeks to optimize work, power, and their reliabilities simultaneously—should avoid fully collective or parallel charging. In fact, strong many-body entanglement, which increases as $\frac{k}{N} \to 1$, can enhance charging or discharging power, it is detrimental to achieving high reliability in power. Instead, one should employ a \emph{hybrid charging} strategy, selecting an intermediate value of $k$ such that $1 < k < N$, thereby balancing the benefits of both parallel and collective regimes.

 The paradigmatic quantum battery models studied above are among the cleanest and most analytically tractable frameworks, making them ideal for demonstrating fundamental trade-off relations, performance bounds, and scaling laws. However, such models are often difficult to realize or faithfully simulate in experimentally accessible platforms. 

In contrast, most experimentally relevant quantum battery models studied in the literature are analytically intractable in the many-body regime, particularly for large numbers of spins or quanta-cells. Their properties can only be investigated using numerical methods. To demonstrate the applicability and universality of our results beyond analytically solvable models, we consider an Ising-type quantum battery, which captures key features of interacting many-body systems that may be realized or effectively simulated in a variety of physical platforms, as we discuss below.

\subsubsection{ Transverse-field Ising-like models with  $s$-body Interactions}
We now consider many-body spin-chain quantum batteries with non-local interactions ranging from short-range to finite long-range in order to examine the figures of merit discussed in the previous sections. We consider the transverse field Ising model with s-body interactions as the charging Hamiltonian of a battery, which is given by
\begin{equation}\label{XX}
H^{I,C}_{s,t} = -\Omega_{s} \sum_{i=1}^{N-s+1} \bigotimes_{j=0}^{s-1} \sigma_x^{(i+j)} - H_0^{B},
\end{equation}
where $2 \leq s \leq N$. The total Hamiltonian is then $H^T=H_0^B + H^{I,C}_{s,t}$. Here, as we increase the value of $s$, the Hamiltonian in Eq.~\eqref{XX} enhances the degree of non-locality of the interactions, thereby interpolating from short-range to long-range couplings. Depending on the $s$ value above generates $s$-body interaction between the spins (see Supplementary Material~\ref{transverse}). It is important to note that, in our analysis, during the charging process we normalized the total Hamiltonian $H^{T}$ such that its eigenvalues lie between 0 and 1, i.e, $H^{T}_{norm}={(H^{T}-E_{min}\mathbb{I}_d)}/{(E_{max}-E_{min})}$, with $E_{max}$ and $E_{min}$ are largest and smallest eigenvalues of $H^{T}$, respectively and $\mathbb{I}_d$ is identity matrix which has same dimension ($d$) as $H^T$. Here, we consider normalized total Hamiltonian to make a fair comparison across the models with different $s$-values. In such cases, the counting operator of power reads as  $P_0= -\frac{i}{\hbar}[H^{B}_0,\tilde{H}^{C}_t]$, where $\tilde{H}^{C}_t={H}^{C}_t/\norm{H^{T}}$. During the charging process of the quantum battery via the Hamiltonian in Eq.~\eqref{XX}, we consider the initial (discharged) state of the battery as $\ket{\psi_0} = \ket{0}^{\otimes N}$, with $N=10$. Here we consider $s = 2$, $3$, and $4$, corresponding to two-, three-, and four-body interactions, respectively, and compute the averages, fluctuations, and NSRs of both work and power. We observe that increasing the value of $s$, which induces longer-range interactions among the spins, enhances the charging power, as shown in Fig.~\ref{XX-Plots}(a). At the same time, this increase in $s$ leads to larger fluctuations and a higher noise-to-signal ratio (NSR) of the power, thereby reducing its reliability, as also illustrated in Fig.~\ref{XX-Plots}(a). Conversely, as displayed in Fig.~\ref{XX-Plots}(b), increasing $s$ results in larger fluctuations in the work; however, the corresponding NSR decreases. Consequently, the reliability of the work improves. For completeness, the plots of work and power fluctuations are provided in Supplementary Material~\ref{transverse}. We also note that as $s$ increases, i.e., as $\frac{s}{N}$ grows for fixed $N$, the average power increases while its reliability decreases. At the same time, the reliability of the work improves. Consequently, charging the quantum battery using the Hamiltonian in Eq.~\eqref{XX} exhibits behavior analogous to that observed when charging via the paradigmatic $k$-body interaction Hamiltonian in Eq.~\eqref{eq:HcManyBody}, as discussed in the previous section. Moreover, as the interactions become increasingly non-local, the product of the NSRs of work and power also increases, as shown in Fig.~\ref{XX-Plots}(c).

The features observed in paradigmatic and Ising-like models suggest that, although strong multipartite entanglement generated during charging or discharging processes can enhance the power of many-body quantum batteries, it is accompanied by a significant reduction in power reliability. Moreover, when the objective is to optimize the reliability of both power and work simultaneously, strong long-range interactions generating large multipartite entanglement generally prove disadvantageous. Instead, a more effective strategy for designing reliable quantum batteries is to employ charging dynamics that exploit interactions of intermediate range.

\section{Conclusions \label{sec:Conclusions}}
Most studies on quantum batteries have primarily focused on enhancing power. However, the practical usefulness of a quantum battery depends on its reliability of its work output and power during charging and discharging processes. Such reliability is naturally quantified by the noise-to-signal ratios (NSRs) of work and power, where high NSR signifies low reliability. In this work, we have established fundamental limitations on the reliability of quantum batteries. First, we derived universal lower bounds on the NSRs of work and power, showing that both are constrained by the function of the charging speed and a temporal correlation function. This reveals an intrinsic limitation on the reliability of work and power of quantum batteries. Second, we demonstrated that quantum mechanics imposes a fundamental trade-off between the reliabilities of work and power, originating from the non-commutativity of their corresponding operators. As a consequence, it is not possible to simultaneously suppress NSRs of both quantities. We found that improving the reliability of power necessarily comes at the expense of reduced reliability in work, and vice versa. We further analyzed these reliability constraints in many-body quantum batteries under parallel (local), collective (fully non-local), and hybrid (semi-local) charging protocols. While collective charging and multipartite interactions can significantly enhance charging power, we found that they also amplify NSR, leading to degraded reliability in power. In contrast, parallel charging exhibits superior reliability in power, particularly for large number of battery cells, while hybrid schemes offer an optimization between power enhancement and reliability in work and power. Similar features are also observed for charging schemes exploiting Ising-like interactions. These results demonstrate that entanglement-assisted high power does not automatically translate into a reliable quantum battery, and that reliability must be treated as an independent and essential figure of merit. Our findings highlight that the optimal design of faithful quantum batteries requires a balanced optimization of power, work, and their reliabilities, rather than maximizing power alone. The fundamental bounds and trade-off relations derived in this work provide guiding principles for identifying such optimal operating regimes.

In the future, the bounds established in this work can be extended to more general models of closed many-body quantum batteries with interactions beyond the paradigmatic and Ising-like spin-chain based models considered here. It would also be interesting to explore how these reliability constraints manifest in photonic quantum batteries and in dissipative or open quantum system charging scenarios, where environmental effects may introduce additional trade-offs. We expect that our results will serve as a foundational step toward designing and realization of high-performance reliable quantum batteries.\\

{\it \bf Acknowledgment} --  BM acknowledges funding by the Research Council of Finland by grant no 355824. TP acknowledges research funding from the QVLS-Q1 consortium, supported by the Volkswagen Foundation and the Ministry for Science and Culture of Lower Saxony and also the Research Council of Finland for funding through Grant No. 359284/Finnish Quantum Flagship. ML acknowledges support from:
MCIN/AEI (PGC2018-0910.13039/501100011033,  CEX2019-000910-S/10.13039/501100011033, Plan National STAMEENA PID2022-139099NB, project funded MCIN and  by the “European Union NextGenerationEU/PRTR" (PRTR-C17.I1), FPI); Ministry for Digital Transformation and of Civil Service of the Spanish Government through the QUANTUM ENIA project call - Quantum Spain project, and by the European Union through the Recovery, Transformation and Resilience Plan - NextGenerationEU within the framework of the Digital Spain 2026 Agenda; CEX2024-001490-S [MICIU/AEI/10.13039/501100011033]; Fundació Cellex;
Fundació Mir-Puig; Generalitat de Catalunya (European Social Fund FEDER and CERCA program; Barcelona Supercomputing Center MareNostrum (FI-2023-3-0024);
Funded by the European Union (HORIZON-CL4-2022-QUANTUM-02-SGA, PASQuanS2.1, 101113690, EU Horizon 2020 FET-OPEN OPTOlogic, Grant No 899794, QU-ATTO, 101168628),  EU Horizon Europe Program (No 101080086 NeQSTGrant Agreement 101080086 — NeQST). BM, TP, and MNB align with the manifesto of Quantum Scientists for Disarmament \cite{Disarmament2026} and oppose any use of the results of this work for military or surveillance purposes. \\

\setcounter{subsection}{0}  
\section{Supplementary Material}

Here, we provide the proof of the bounds and detailed calculations which supplement the main text.

\subsection{ Average and Fluctuation of Work and Power using Full Counting Statistics}~~\label{FCS}
In this section, we first discuss the relationship between full counting statistics (FCS) and the two-point measurement protocol (TPM) for measuring a quantum observable (for more details see Ref.~\cite{Esposito2009}). We consider an counting observable \( O_0 \) to account the corresponding physical quantity during physical process.  In general, one can consider  explicit time dependent observable \( O(t) \), however we do not need in our case.
Let us consider a quantum system whose time evolution is governed by unitary dynamics, and its time-evolved state at time t is given by $\rho_t=U_t\rho_0U_t^{\dagger}$, where $\rho_0$ and $\rho_t$ are initial and time evolved state of the system respectively, and $U=e^{-iHt/\hbar}$ and H is driving Hamiltonian of the system. We want to measure the observable \( O_0 \) at each instant of time during the evolution. To do this, we first parametrized the time-evolved density matrix \( \rho(\chi,t) \) by a counting field \( \chi \), such as:
\begin{equation}
\label{EOM1}
\rho(\chi,t) = U(\chi,t) \rho_0 U^{\dagger}(-\chi,t),
\end{equation}
where \( U(\chi,t) \) describes the modified unitary evolution due to the introduction of the counting field:
\(U(\chi,t) = e^{i(\chi O_0/2)} U_t e^{-i(\chi O_0/2)}, \ \mbox{and} \ 
U^{\dagger}(-\chi,t) = e^{-i(\chi O_0/2)} U^{\dagger}_t e^{i(\chi O_0/2)}
 \). The generating function for the full counting statistics is encoded in the partition function \( Z(\chi,t) \), which is obtained by tracing over the modified density matrix \( \rho(\chi,t) \):
\begin{align}
Z(\chi,t) &= \ln{\Tr(\rho(\chi,t))}\nonumber \\
 &= \ln{\Tr\left( e^{-i\chi O_0/2} e^{i\chi O^H_t} e^{-i\chi O_0/2} \rho_0 \right)},
\end{align}
where \( O_t = U^{\dagger}_t O_0 U_t \) is the time evolved counting observable in the Heisenberg picture. The average value of quantity associated with the counting observable \( O_0 \) at time $t$ can be obtained from the generating function as:
\begin{align}
 \langle  \cal{O}_t \rangle &= \frac{d}{d(i\chi)} Z(\chi,t)|_{\chi=0}= \Tr(\rho_0\cal{O}_t),   
\end{align}
where we defined $\cal{O}_t=O_t-O_0$. Additionally, the variance of quantity associated with counting observable \( O_0 \) of the observable can be calculated from the second derivative of the generating function:
\begin{align}
(\Delta \mathcal{O}_t)^2
&= \left(\frac{d}{d(i\chi)}\right)^2 Z(\chi,t)\Big|_{\chi=0} \nonumber \\
&= \tr\!\left(\mathcal{O}_t^2 \rho_0\right)
   - \tr\!\left(\mathcal{O}_t \rho_0\right)^2 .
\end{align}

This formulation establishes the connection between FCS and TPM and provides insights into the fluctuations of the observable over time by considering its variance~\cite{Esposito2009}.
In the context of quantum batteries, if the counting observable is internal energy $W_0=H^{B}_0$ the above definitions provide the average work and fluctuation in work in the charging process, given as
\begin{equation}
\langle \cal{W}_t \rangle = \Tr(\rho_0 \cal{W}_t ) \ \mbox{and} \   (\Delta  \cal{W}_t)^2   =\tr(\cal{W}_t^2\rho_0) -\tr(\cal{W}_t\rho_0)^2,
\end{equation}
where $\cal{W}_{t}= H^{B}_{t}-H^{B}_{0}$ can be regarded as the Heisenberg operator of work \cite{Allahverdyan2005}.
Important to note that average work is same  with the definition of ergotropy~\cite{Allahverdyan2004}. The above expression of average and fluctuation in work aligns with TPM scheme, if initial state commutes with $H^{B}_0$. It is important to note that the TPM scheme fails to account for the correct
energy change when the initial state possesses coherence in the energy basis,
since it does not commute with \( H_0^{B} \) \cite{Allahverdyan2005, Talkner2016}. In contrast, the expressions obtained using full counting statistics (FCS)
naturally incorporate initial energy coherence see Refs.~\cite{Solinas2015,Xu2018, Francica2022, Francica2023}, remain valid beyond the TPM scheme.
The flux of energy $\frac{d\langle H_{B} \rangle}{dt}=-\frac{i}{\hbar}\tr([H^{B}_0,H^{C}_t]\rho_{t})$, which let us define the counting observable $P_0= -\frac{i}{\hbar}[H^{B}_0,H^{C}_t]$, then we obtain the average power and fluctuation
in power in the charging process, given as
\begin{equation}
\langle \cal{P}_t \rangle = \Tr(\rho_0 \cal{P}_t ) \ \mbox{and} \   (\Delta  \cal{P}_t)^2   =\tr(\cal{P}_t^2\rho_0) -\tr(\cal{P}_t\rho_0)^2
\end{equation}
Note that here, symbols $\cal{W}$ and $\cal{P}$ stands for work and power, respectively. It is worthwhile to mention that, in general, the averages and fluctuations of work and power are distinct from the average and fluctuations (variances) of the instantaneous energy and its rate at an arbitrary time $t$. It is important to mention here that throughout this work, we considered normalized total Hamiltonian for fair comparison across the model, therefore in such case the counting operator of power reads as  $P_0= -\frac{i}{\hbar}[H^{B}_0,\tilde{H}^{C}_t]$, where $\tilde{H}^{C}_t={H}^{C}_t/\norm{H^{T}}$.

\subsection{ Proof of Fundamental Bounds on Reliabilities}\label{PLB}
Let \( O \) be an observable with spectral decomposition  
\( 
O = \sum_j o_j |j\rangle\langle j| 
\). Consider a quantum system undergoing unitary evolution, with its state at time \( t \) denoted by \( |\psi_t\rangle \). At two different times, \( \tau_1 \) and \( \tau_2 \), the probability distributions associated with the measurement of observable \( O \) are given by  \(
p_j = |\langle j | \psi_{\tau_1} \rangle|^2\) and \(, \quad q_j = |\langle j | \psi_{\tau_2} \rangle|^2
\). We denote these distributions by \( P = \{p_j\} \) and \( Q = \{q_j\} \).  
The distance between \( P \) and \( Q \) can be quantified using the Hellinger distance, defined as  
\begin{equation}\label{HD}
H^2(P, Q) = 1 - \sum_j \sqrt{p_j q_j}.
\end{equation}  
The Hellinger distance is bounded from below as (see Refs.~\cite{Nishiyama2021, Nishiyama2025} ) 
\begin{equation}\label{HDNSR}
H^2(P, Q) \geq 1 - \left[ \left( \frac{\mu_P - \mu_Q}{\sigma_P + \sigma_Q} \right)^2 + 1 \right]^{-1/2},
\end{equation}
where the expectation values and standard deviations of \( O \) at times \( \tau_1 \) and \( \tau_2 \) are defined as  
\(
\mu_X = \langle O \rangle(\tau_i), \quad 
\sigma_X = \Delta O(\tau_i)
\).  Thus, combining above Eqs.~\eqref{HD} and \eqref{HDNSR}, we obtain  
\begin{equation}\label{Hasegawa}
 \left[ \left( \frac{\langle O \rangle(\tau_1) - \langle O \rangle(\tau_2)}
 {\Delta O(\tau_1) + \Delta O(\tau_2)} \right)^2 + 1 \right]^{-\frac{1}{2}} \geq \sum_j  \sqrt{p_j(\tau_2)\, q_j(\tau_1)}.
\end{equation}

The evolution time is \( t = \tau_2 - \tau_1 \). Using the Bhattacharyya bound~\cite{Hasegawa2023}, and assuming \( \tau_1 = 0 \) and \( \tau_2 = t \), we obtain  
\begin{equation}\label{BB}
 \int_{0}^{t} \sqrt{\frac{\mathcal{I}^O_{t'}}{4}} \, dt' 
 \;\; \geq \;\; \arccos\!\left( \sum_{i}\sqrt{p_i(t)\,p_i(0)} \right) ,
\end{equation}
where $\mathcal{I}^{O}_{t}$ is the classical Fisher information in the eigenbasis of observable \( O \), defined as  
\begin{equation}\label{FI}
    \mathcal{I}^{O}_{t} = \sum_{i}\frac{1}{p_{i}(t)}\left(\frac{dp_{i}(t)}{dt}\right)^{2},
\end{equation}
and the probabilities \( p_i(t) = \mathrm{Tr}(\rho_t \Pi_i ) \) correspond to the projectors \( \Pi_i =\op{i}\) of the observable  
\( O = \sum_{i} o_i \Pi_i\) in the time-evolved state \( \rho_t \).  Combining Eqs.~\eqref{Hasegawa} and \eqref{BB}, we obtain  
\begin{equation}\label{NSRB}
\left[ \frac{\Delta O_t + \Delta O_0}{\langle O_t \rangle - \langle O_0 \rangle} \right]^{2}
\geq \frac{1}{\tan^{2}\!\left(\int_{0}^{t} \sqrt{\tfrac{\mathcal{I}^{O}_{t'}}{4}} \, dt'\right)}
.
\end{equation}  
Note that in the above step we also switch from the Schrödinger picture to the Heisenberg picture, i.e., transferring the time dependence from the state to the observable. Now, let us take \( O = O_0 \) to be a counting observable. The noise-to-signal ratio (NSR) of the corresponding measured quantity is defined as  
\begin{equation}\label{NSRD}
\cal{N}_t^{\cal{O}} =\frac{(\Delta \cal{O}_t)^2 }{\langle \cal{O}_t  \rangle^2} = \frac{(\Delta {O}_t)^2 +  (\Delta {O}_0)^2 - 2 cov(O_t, O_0)}{\langle ({O}_t  - {O}_0)\rangle^2}
\end{equation}
where \(
\operatorname{cov}(O_t, O_0) = \tfrac{1}{2} \langle \{ O_t, O_0 \} \rangle - \langle O_t \rangle \langle O_0 \rangle\) with
 \(\{A, B\} = AB + BA \). Using Eqs.~\eqref{NSRB} and \eqref{NSRD}, we arrive at the bound  
\begin{equation}\label{NSRfinal}
\mathcal{N}_t^{\cal{O}} \;\; \geq \;\; 
{\cot^{2}\!\left(\int_{0}^{t} {\tfrac{\sqrt{\mathcal{I}^{O_0}_{t'}}}{2}} \, dt'\right)} \; - \; f(O_0,O_t)
,
\end{equation}
where the last term  in the left hand side 
\[
f(O_0,O_t) = 2\,\frac{\mathrm{cov}(O_t,O_0) + \Delta O_t \Delta O_0}{\langle O_t \rangle^2}
\]
encodes the temporal correlation between \( O_0 \) and \( O_t \), and vanishes if the system is in an eigenstate of either of them.
This completes the proof of bound given in Eq.~\eqref{NSRBB} of the main text.

\subsection{ Proof of the Fundamental trade-off between Reliabilities}~~\label{FTFR}
Let us recall the famous Robertson-Schrödinger uncertainty relation~\cite{Robertson1929, Schrodinger1930} states that for any two observables \( A \) and \( B \), we have:
\begin{equation}
    (\Delta A)^2 (\Delta B)^2 \geq \frac{1}{4} \left| \langle [A, B] \rangle \right|^2 + \frac{1}{4} \left| \langle \{ A, B \} \rangle - 2 \langle A \rangle \langle B \rangle \right|^2.
\end{equation}

For the given case, we identify the Hermitian operators:
\(
A = \tilde{\mathcal{P}} = \frac{\mathcal{P}}{\langle \mathcal{P}_t \rangle}\) and \( \quad B =\tilde{\mathcal{W}} = \frac{\mathcal{W}}{\langle \mathcal{W}_t \rangle}
\). Using the expectation values taken over the initial state \( \rho_0 \), we set:
\(
\langle A \rangle = \text{Tr}(\tilde{\cal{P}} \rho_0)\) and \(\quad \langle B \rangle = \text{Tr}(\tilde{\cal{W}} \rho_0)
\). From the uncertainty relation, we obtain \((\Delta \tilde{\mathcal{P}})^2 \, (\Delta \tilde{\mathcal{W}})^2 \geq 
\; \frac{1}{4} \left| \text{Tr}\!\left([\tilde{\mathcal{P}}, \tilde{\mathcal{W}}] \rho_0 \right) \right|^2 \nonumber 
+ \frac{1}{4} \left| \text{Tr}\!\left(\{\tilde{\mathcal{P}}, \tilde{\mathcal{W}}\} \rho_0 \right) 
- 2 \langle \tilde{\mathcal{P}} \rangle \langle \tilde{\mathcal{W}} \rangle \right|^2 \)
 . Since the noise-to-signal ratio is defined in a way that normalizes expectation values, we make use of the fact that: \(
\langle \tilde{\cal{P}} \rangle \langle \tilde{\cal{W}} \rangle = 1
\). Thus, the second term simplifies to: \(\left| \frac{1}{2} \text{Tr}(\{\tilde{\cal{P}},\tilde{\cal{W}}\} \rho_0) - 1 \right|^2
\). Therefore, we arrive at the required bound~\eqref{TOWP} in main text. This completes the proof.

\subsection{Estimation of Bounds for Single Qubit Battery}~\label{SQB}

Let us consider a model qubit battery with internal Hamiltonian $H^{B}_{0} = -\frac{\omega_0}{2} \sigma_z$ and a charging Hamiltonian given by $H^{C}_{t} = \Omega_0 \sigma_x - H^{B}_{0}$ with $t \in [0,T]$, where $\sigma_z$ and $\sigma_x$ denote the Pauli $z$ and $x$ matrices, respectively, and $ \Omega_0 \in \mathbb{R}$. The counting observables for work and power are given as (assuming $\hbar=1$)
\begin{equation}
 W_0  = -\frac{\omega_0}{2}\sigma_{z}, \ \mbox{and} \  P_0  = -i[H^{B}_0,H^{C}_t]=-\Omega_0\omega_{0}\sigma_{y}.
\end{equation}
The evolution operator in this case is given as 
\begin{equation}
    U_{t}= e^{-it H^T } = e^{-i \Omega_0t \sigma_x} = \cos(\Omega_0t)\, \mathbb{I}_{2} - i \sin(\Omega_0t)\, \sigma_x.
\end{equation}
The time evolved counting observable of work and power operator are
\[
\begin{aligned}
W_t &= e^{it H^T } W_0 e^{-it H^T } \\
    &= -\frac{\omega_0}{2} \left[ \cos(2\Omega_0 t)\, \sigma_z + \sin(2\Omega_0 t)\, \sigma_y \right].
\end{aligned}
\]

and

\[
\begin{aligned}
P_t &= e^{it H^T } P_0 e^{-it H^T } \\
    &= -\Omega_0 \omega_0 \left[ \cos(2\Omega_0 t)\, \sigma_y - \sin(2\Omega_0 t)\, \sigma_z \right].
\end{aligned}
\]

If we assume the initial $|\psi_0\rangle= |0\rangle$ then average and variance of work and power given as
\begin{align*}
\langle \cal{W}_t \rangle = \omega_{0} \sin ^2(\Omega_0t),  (\Delta \cal{W}_t)^2 = \frac{\omega_{0}^{2}}{4} \sin ^2(2\Omega_0t) ,
\end{align*}
and 
\begin{align*}
&\langle \cal{P}_t \rangle = \Omega_0\omega_{0} \sin (2\Omega_0t) ,(\Delta \cal{P}_t )^2 = 4 \Omega_0^{2} \omega_{0}^{2} \sin ^4(\Omega_0t).
\end{align*}
The noise-to-signal ratios of work and power are given as
\begin{equation}
\frac{(\Delta \cal{W}_t)^2}{\langle \cal{W}_t \rangle^2 } =  \cot^2{(\Omega_0t)}, \ \mbox{and} \frac{(\Delta \cal{P}_t)^2}{\langle \cal{P}_t \rangle^2 } =  \tan^2{(\Omega_0t)}.
\end{equation}
The product of the NSRs of work and power can be obtained 
\begin{equation}
\frac{(\Delta \cal{W}_t)^2}{\langle \cal{W}_t \rangle^2 }\frac{(\Delta \cal{P}_t)^2}{\langle \cal{P}_t \rangle^2 } = 1.
\end{equation}
Important to note that, since \( 
\tr([\cal{P}_t,\cal{W}_t]\rho_0) =0  \) and \(\tr(\{\cal{P}_t,\cal{W}_t\}\rho_0) =0 \). Hence, it saturates the trade off relation between NSRs of work and power given by Eq.~\eqref{TOWP}.
Moreover $f(W_0,W_t)=0$, \( 
f(P_0, P_t) = 4 \cot(2 \Omega_0 t) \csc(2 \Omega_0 t)
\) and \( \mathcal{I}^{P_0/W_0}_{t} = 4 \Omega_0^2 \). 
The lower bounds on the NSRs of work and power are 
$\cot^2(\Omega_0 t)$ and $\cot^2(\Omega_0 t) - 4 \cot(2\Omega_0 t) \csc(2\Omega_0 t) = \tan^2(\Omega_0 t)
$, respectively. 
Thus, the bound given in Eq.~\eqref{NSRBB} saturates for  both work and power.

\begin{figure*}
\centering
\includegraphics[width=0.33\textwidth]{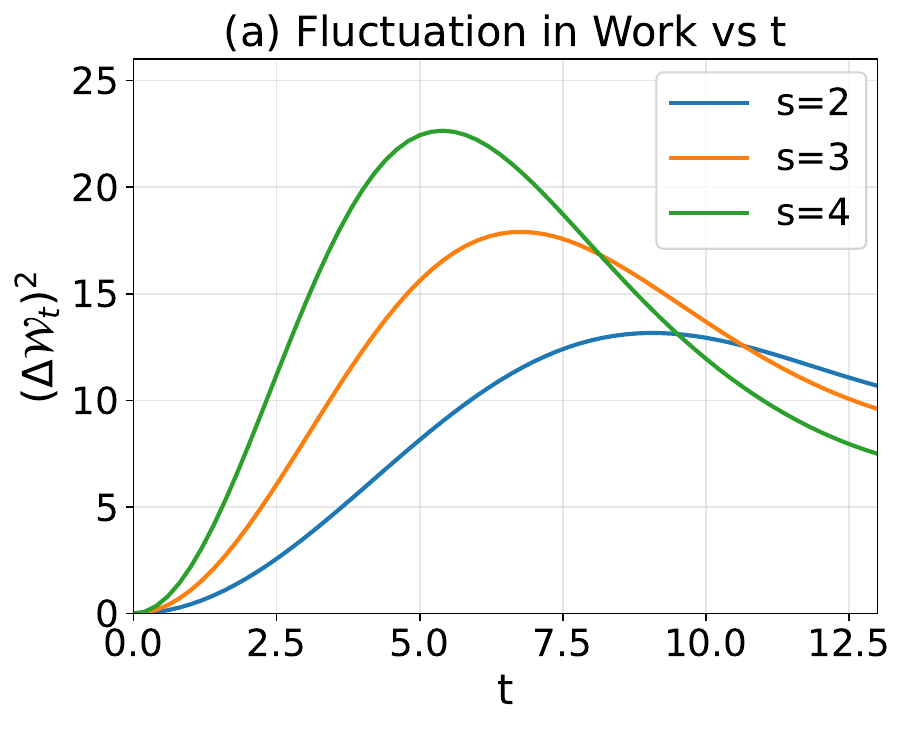}
\includegraphics[width=0.33\textwidth]{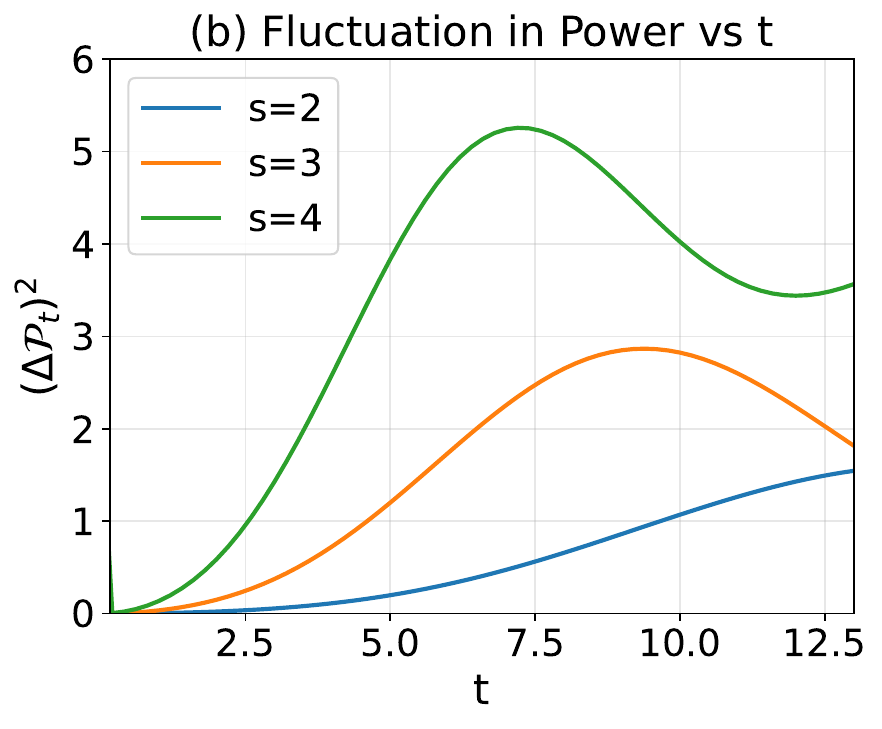}
\includegraphics[width=0.33\textwidth]{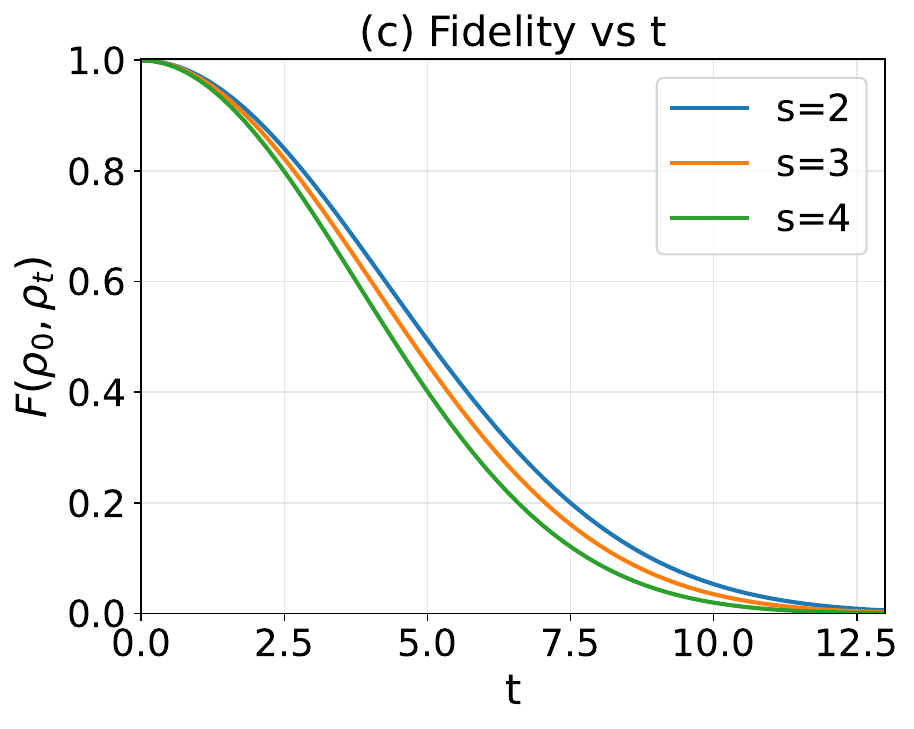}
\caption{\justifying{ The plots (a)–(c) supplements the Fig.~~\ref{XX-Plots} in the main text. Here, we depict (in plots (a)–(c)) that fluctuations in work and power along with fidelity (between initial and time evolved state) during charging process. Here, we have considered, total spin $N=10$, $\Omega_s = 1$ and $\omega_0=2.$}}
\label{XX-flu}
\end{figure*}

\subsection{ Charging Dynamics of Quantum Battery with $k$-Party Interactions}
\label{sec:k body hamiltonian}
We analyze the dynamics of a quantum battery composed of \( N = k q \) qubits, divided into \( q \) disjoint blocks of \( k \) qubits each, with initial state \( |0 \rangle^{\otimes N} \). The bare Hamiltonian is \( H^B_0 = -\frac{\omega_0}{2} \sum_{j=1}^{N} \sigma_z^{(j)} = \sum_{j=0}^{q-1}  H^{B,j}_0  \), where \( H^{B,j}_0 = -\frac{\omega_0}{2} \sum_{i=1}^k \sigma_z^{(k j + i)} \). The charging Hamiltonian can be written as
\[
H_{k,t}^C = \Omega_k \sum_{j=0}^{q-1} X_j- H_{0}^{B} , \quad X_j = \bigotimes_{i=1}^k \sigma_x^{(k j + i)} ,
\]
with commuting block operators \( X_j \). The time-evolution unitary is
\[
U_t = e^{-i t H^T} = \prod_{j=0}^{q-1} \left[ \cos(\Omega_k t) \mathbb{I}_k - i \sin(\Omega_k t) X_j \right].
\]

 Since, each block evolves independently in the two-dimensional subspace spanned by \( |g\rangle = |0 \rangle^{\otimes k} \) and \( |e\rangle = |1 \rangle^{\otimes k} \), we can map it to an effective two-level system (qubit system) with \( X_j \to \sigma_x^\text{eff} \), \( |g\rangle \to |0\rangle_\text{eff} \), \( |e\rangle \to |1\rangle_\text{eff} \), and, for the initial state, \( \langle \sigma_z^\text{eff} \rangle = 1 \), \( \langle \sigma_y^\text{eff} \rangle = 0 \). The counting observable for power is 
\begin{align}
P_0 &= -i [H^B_0, H_{k,t}^C] = \sum_j P_0^j \nonumber \\
&= -\Omega_k \omega_0 \sum_{j=0}^{q-1} \sum_{l=1}^k 
\sigma_y^{(k j + l)} \bigotimes_{\substack{i=1 \\ i \neq l}}^k 
\sigma_x^{(k j + i)}.
\end{align}
The time evolved power and work counting observables are \( P_t = U_t^\dagger P_0 U_t = \sum_j P_t^j \), and \( W_{t} = U_t^\dagger H_{0}^B U_t = \sum_j W_{t}^j \). For the $j$-th block, the effective Hamiltonian is $ H^{B,j}_0 = -(\omega_0 k/2)\, \sigma_z^\text{eff}$, which evolves as $\sigma_z^\text{eff}(t) = \cos(2 \Omega_k t)\, \sigma_z^\text{eff} + \sin(2 \Omega_k t)\, \sigma_y^\text{eff}$, yields work operator at time $t$ given as
\begin{equation}
\begin{aligned}
\mathcal{W}_t^j
&= W_t^j-W_0^j \\
&= \frac{\omega_0 k}{2}
\left[
\left(1-\cos(2\Omega_k t)\right)\sigma_z^{\mathrm{eff}}
-\sin(2\Omega_k t)\,\sigma_y^{\mathrm{eff}}
\right].
\end{aligned}
\end{equation}
Now, we calculate the average work and fluctuation in work during charging at an arbitrary time $t$ by summing over all $q$ blocks, given as
\begin{align}\label{k-Work}
\langle \cal{W}_t \rangle &= N \, \omega_0 \sin^2(\Omega_k t), \ \mbox{and} \\
(\Delta \cal{W}_t)^2 &= \frac{N k \omega_0^2}{4} \sin^2(2 \Omega_k t).
\end{align}
The power operator of $j$-th block at time $t$ is given as
\begin{equation}\label{k-Power}
\begin{aligned}
\cal{P}_t^j &= P_t^j - P_0^j\\ 
&= \Omega_k \omega_0 k 
\big[ (1-\cos 2\Omega_k t)\, \sigma_y^{\text{eff}} 
+ \sin 2\Omega_k t\, \sigma_z^{\text{eff}} \big].
\end{aligned}
\end{equation}
After doing some algebra, we obtain the average power and fluctuation in power during charging at an arbitrary time $t$ by summing over all $q$ blocks, given as
\begin{align}
\langle \mathcal{P}_t \rangle &= N\,\Omega_k \omega_0 \sin(2\Omega_k t),\\
(\Delta \mathcal{P}_t)^2 &= 4 N k\,\Omega_k^2 \omega_0^2 \sin^4(\Omega_k t).
\end{align}

The product of NSRs of work and power in parallel and collective charging of multi-qubit batteries are given as
\begin{equation}\label{eq:NSRproduct}
\begin{aligned}
\mathcal{N}^{\mathcal{W},(k=1)}_{t}\mathcal{N}^{\mathcal{P},(k=1)}_{t} &= \frac{1}{N^{2}},\\
\mathcal{N}^{\mathcal{W},(k=N)}_{t}\mathcal{N}^{\mathcal{P},(k=N)}_{t} &= 1.
\end{aligned}
\end{equation}
This provides a clear distinction between parallel and collective charging: in the collective case, the product is fixed at unity, whereas in the parallel case, it is suppressed by a factor of $1/N^2$. For hybrid case, consider the case, when $N/k>>\Omega_0t$, we can expand the trigonometric functions as $\sin\theta \approx \theta$, $\cos\theta \approx 1$, and $\tan\theta \approx \theta$.  
Using the above Eqs. and approximations, we have (for scenario $N/{k}>>\Omega_0 t$)
\begin{align}
\mathcal{N}^{\mathcal{W},(k)}_{t} = \frac{{k}}{N} \cot^2 \!\left(\frac{{k}t}{N}\Omega_0\right) \approx \frac{N}{{k}}\frac{1}{\Omega_0^2t^2}, \label{eq:NSRlargeW}
\end{align}
while from Eq.~\eqref{k-Power}, we obtain
\begin{align}
\mathcal{N}^{\mathcal{P},(k)}_{t} = \frac{{k}}{N} \tan^2 \!\left(\frac{{k}t}{N}\Omega_0\right) \approx \frac{{k^3}}{N^3}{\Omega_0^2t^2}. \label{eq:NSRlargeP}
\end{align}
A quick multiplication, suggest that in the considered limit the product of NSRs of hybrid case remains \(\mathcal{N}^{\mathcal{W},(k)}_{t}\mathcal{N}^{\mathcal{P},(k)}_{t} = \frac{k^2}{N^{2}}\). Thus, we retrieve the universal cluster scaling given Eq.~~\eqref{eq:NSRScaling} of main text. Moreover, here the proportionality relations of NSRs with time also 
corroborated with plot in Fig-~~\ref{fig:kbodyplot} (c).

\

{\it \bf Scaling of NSRs in Hybrid Charging via k-body Hamiltonian for $n=\frac{N}{k} \notin \mathbb{N}$ case:} Let us consider the intermediate charging case, i.e., charging with k-body charging Hamiltonian $H_k^C$ with $\norm{H_{1}^T} =  \norm{H_{k}^T}$. In this case counting observable for work is a separable operator like previous cases, however, the counting observable for power is a k-body operator. Therefore, in this charging schemes, if we ignore the trigonometric functions then that maximum scaling of work and power are  \(\langle \mathcal{W} \rangle \sim nk\) \ \mbox{and} \ \( \langle \mathcal{P} \rangle \sim nk \Omega_{k}
\), and their maximum fluctuations are obtained using $\Omega_{k} \geq \Omega_{N-nk}$,
\begin{equation}
\begin{aligned}
(\Delta \mathcal{W})^2
&\sim nk^2 + (N-nk)^2, \\
(\Delta \mathcal{P})^2
&\sim \Omega_k^2\left[nk^2 + (N-nk)^2\right].
\end{aligned}
\end{equation}
where N-total Spin and assuming $n=N/k$ is not an integer, i.e., $n=\frac{N}{k} \notin \mathbb{N}$. Thus the scaling of product of NSRs given as
\begin{equation}
    \mathcal{N}^{\mathcal{W}} \mathcal{N}^{\mathcal{P}} \sim\frac{(  nk^2+ (N-nk)^2)^{2}}{(n^{2}k^{2})^2}
\end{equation}
For the non-integer case considered above, for $N/k \notin \mathbb{N}$, we take $n=\left\lfloor N/k \right\rfloor$, where $n$ denotes the number of complete $k$-spin blocks, and $N-nk$ denotes the number of remaining spins. Important to note that, if $n=\frac{N}{k} \in \mathbb{N}$ is integer the term $(N-nk)^2$ vanishes in the above scalings. Thus the scaling of product of NSRs reduces to
\begin{equation}
    \mathcal{N}^{\mathcal{W}} \mathcal{N}^{\mathcal{P}}  \sim \frac{k^{2}}{N^{2}}.
\end{equation}

Likewise the single qubit case, it is straightforward to analysis to check the bound~\eqref{TOWP} saturates for parallel, collective as well as hybrid charging case.

\subsection{Charging QBs via Transverse Field like Ising model with  $s$-body Interactions}~\label{transverse}
The charging Hamiltonian written in the Eq.~\eqref{XX}, for $s=2,3,4$ generates 2-body, 3-body and 4-body interactions respectively. The charging Hamiltonian Eq.~\eqref{XX} for $N=10$, spins is given as (for $s=2,3,4$)
\begin{align*}
H^{I,C}_{2,t} &= -\Omega_{s} \sum_{i=1}^{9} \sigma_i^x \sigma_{i+1}^x 
                + \frac{\omega_{0}}{2} \sum_{i=1}^{10} \sigma_i^z, 
\end{align*}

\begin{align*}
H^{I,C}_{3,t} &= -\Omega_{s} \sum_{i=1}^{8} \sigma_i^x \sigma_{i+1}^x \sigma_{i+2}^x 
                + \frac{\omega_{0}}{2} \sum_{i=1}^{10} \sigma_i^z,
\end{align*}
and

\begin{align*}
H^{I,C}_{4,t} &= -\Omega_{s} \sum_{i=1}^{7} \sigma_i^x \sigma_{i+1}^x \sigma_{i+2}^x \sigma_{i+3}^x 
                + \frac{\omega_{0}}{2} \sum_{i=1}^{10} \sigma_i^z.
\end{align*}

The Fig~\ref{XX-flu} supplements the Fig.~\ref{XX-Plots} in the main text. In Fig~\ref{XX-flu}, we depict that fluctuations in work and power along with fidelity of initial and time evolved state during charging process. 

 \bibliography{name}

\begin{thebibliography}{72}%
\makeatletter
\providecommand \@ifxundefined [1]{%
 \@ifx{#1\undefined}
}%
\providecommand \@ifnum [1]{%
 \ifnum #1\expandafter \@firstoftwo
 \else \expandafter \@secondoftwo
 \fi
}%
\providecommand \@ifx [1]{%
 \ifx #1\expandafter \@firstoftwo
 \else \expandafter \@secondoftwo
 \fi
}%
\providecommand \natexlab [1]{#1}%
\providecommand \enquote  [1]{``#1''}%
\providecommand \bibnamefont  [1]{#1}%
\providecommand \bibfnamefont [1]{#1}%
\providecommand \citenamefont [1]{#1}%
\providecommand \href@noop [0]{\@secondoftwo}%
\providecommand \href [0]{\begingroup \@sanitize@url \@href}%
\providecommand \@href[1]{\@@startlink{#1}\@@href}%
\providecommand \@@href[1]{\endgroup#1\@@endlink}%
\providecommand \@sanitize@url [0]{\catcode `\\12\catcode `\$12\catcode `\&12\catcode `\#12\catcode `\^12\catcode `\_12\catcode `\%12\relax}%
\providecommand \@@startlink[1]{}%
\providecommand \@@endlink[0]{}%
\providecommand \url  [0]{\begingroup\@sanitize@url \@url }%
\providecommand \@url [1]{\endgroup\@href {#1}{\urlprefix }}%
\providecommand \urlprefix  [0]{URL }%
\providecommand \Eprint [0]{\href }%
\providecommand \doibase [0]{https://doi.org/}%
\providecommand \selectlanguage [0]{\@gobble}%
\providecommand \bibinfo  [0]{\@secondoftwo}%
\providecommand \bibfield  [0]{\@secondoftwo}%
\providecommand \translation [1]{[#1]}%
\providecommand \BibitemOpen [0]{}%
\providecommand \bibitemStop [0]{}%
\providecommand \bibitemNoStop [0]{.\EOS\space}%
\providecommand \EOS [0]{\spacefactor3000\relax}%
\providecommand \BibitemShut  [1]{\csname bibitem#1\endcsname}%
\let\auto@bib@innerbib\@empty
\bibitem [{\citenamefont {Alicki}\ and\ \citenamefont {Fannes}(2013)}]{Alicki2013}%
  \BibitemOpen
  \bibfield  {author} {\bibinfo {author} {\bibfnamefont {R.}~\bibnamefont {Alicki}}\ and\ \bibinfo {author} {\bibfnamefont {M.}~\bibnamefont {Fannes}},\ }\bibfield  {title} {\bibinfo {title} {Entanglement boost for extractable work from ensembles of quantum batteries},\ }\href {https://doi.org/10.1103/PhysRevE.87.042123} {\bibfield  {journal} {\bibinfo  {journal} {Physical Review E}\ }\textbf {\bibinfo {volume} {87}},\ \bibinfo {pages} {042123} (\bibinfo {year} {2013})}\BibitemShut {NoStop}%
\bibitem [{\citenamefont {Binder}\ \emph {et~al.}(2018)\citenamefont {Binder}, \citenamefont {Correa}, \citenamefont {Gogolin}, \citenamefont {Anders},\ and\ \citenamefont {Adesso}}]{Binder2018}%
  \BibitemOpen
  \bibfield  {author} {\bibinfo {author} {\bibfnamefont {F.}~\bibnamefont {Binder}}, \bibinfo {author} {\bibfnamefont {L.~A.}\ \bibnamefont {Correa}}, \bibinfo {author} {\bibfnamefont {C.}~\bibnamefont {Gogolin}}, \bibinfo {author} {\bibfnamefont {J.}~\bibnamefont {Anders}},\ and\ \bibinfo {author} {\bibfnamefont {G.}~\bibnamefont {Adesso}},\ }\href {https://doi.org/10.1007/978-3-319-99046-0} {\emph {\bibinfo {title} {Thermodynamics in the Quantum Regime}}},\ Vol.\ \bibinfo {volume} {195}\ (\bibinfo  {publisher} {Springer International Publishing},\ \bibinfo {year} {2018})\BibitemShut {NoStop}%
\bibitem [{\citenamefont {Auff\`eves}(2022)}]{Alexia2022}%
  \BibitemOpen
  \bibfield  {author} {\bibinfo {author} {\bibfnamefont {A.}~\bibnamefont {Auff\`eves}},\ }\bibfield  {title} {\bibinfo {title} {Quantum technologies need a quantum energy initiative},\ }\href {https://doi.org/10.1103/PRXQuantum.3.020101} {\bibfield  {journal} {\bibinfo  {journal} {PRX Quantum}\ }\textbf {\bibinfo {volume} {3}},\ \bibinfo {pages} {020101} (\bibinfo {year} {2022})}\BibitemShut {NoStop}%
\bibitem [{\citenamefont {Quach}\ \emph {et~al.}(2023)\citenamefont {Quach}, \citenamefont {Cerullo},\ and\ \citenamefont {Virgili}}]{Quach2023}%
  \BibitemOpen
  \bibfield  {author} {\bibinfo {author} {\bibfnamefont {J.~Q.}\ \bibnamefont {Quach}}, \bibinfo {author} {\bibfnamefont {G.}~\bibnamefont {Cerullo}},\ and\ \bibinfo {author} {\bibfnamefont {T.}~\bibnamefont {Virgili}},\ }\bibfield  {title} {\bibinfo {title} {Quantum batteries: The future of energy storage?},\ }\href {https://doi.org/10.1016/j.joule.2023.09.003} {\bibfield  {journal} {\bibinfo  {journal} {Joule}\ }\textbf {\bibinfo {volume} {7}},\ \bibinfo {pages} {2195} (\bibinfo {year} {2023})}\BibitemShut {NoStop}%
\bibitem [{\citenamefont {Kurman}\ \emph {et~al.}(2026)\citenamefont {Kurman}, \citenamefont {Hymas}, \citenamefont {Fedorov}, \citenamefont {Munro},\ and\ \citenamefont {Quach}}]{Kurman2025}%
  \BibitemOpen
  \bibfield  {author} {\bibinfo {author} {\bibfnamefont {Y.}~\bibnamefont {Kurman}}, \bibinfo {author} {\bibfnamefont {K.}~\bibnamefont {Hymas}}, \bibinfo {author} {\bibfnamefont {A.}~\bibnamefont {Fedorov}}, \bibinfo {author} {\bibfnamefont {W.~J.}\ \bibnamefont {Munro}},\ and\ \bibinfo {author} {\bibfnamefont {J.}~\bibnamefont {Quach}},\ }\bibfield  {title} {\bibinfo {title} {Powering quantum computation with quantum batteries},\ }\href {https://doi.org/10.1103/l39v-jwwz} {\bibfield  {journal} {\bibinfo  {journal} {Phys. Rev. X}\ }\textbf {\bibinfo {volume} {16}},\ \bibinfo {pages} {011016} (\bibinfo {year} {2026})}\BibitemShut {NoStop}%
\bibitem [{\citenamefont {Ferraro}\ \emph {et~al.}(2026)\citenamefont {Ferraro}, \citenamefont {Cavaliere}, \citenamefont {Genoni}, \citenamefont {Benenti},\ and\ \citenamefont {Sassetti}}]{Ferraro2026}%
  \BibitemOpen
  \bibfield  {author} {\bibinfo {author} {\bibfnamefont {D.}~\bibnamefont {Ferraro}}, \bibinfo {author} {\bibfnamefont {F.}~\bibnamefont {Cavaliere}}, \bibinfo {author} {\bibfnamefont {M.~G.}\ \bibnamefont {Genoni}}, \bibinfo {author} {\bibfnamefont {G.}~\bibnamefont {Benenti}},\ and\ \bibinfo {author} {\bibfnamefont {M.}~\bibnamefont {Sassetti}},\ }\bibfield  {title} {\bibinfo {title} {Opportunities and challenges of quantum batteries},\ }\href {https://doi.org/10.1038/s42254-025-00906-5} {\bibfield  {journal} {\bibinfo  {journal} {Nature Reviews Physics}\ }\textbf {\bibinfo {volume} {8}},\ \bibinfo {pages} {115} (\bibinfo {year} {2026})}\BibitemShut {NoStop}%
\bibitem [{\citenamefont {Campaioli}\ \emph {et~al.}(2017)\citenamefont {Campaioli}, \citenamefont {Pollock}, \citenamefont {Binder}, \citenamefont {C\'eleri}, \citenamefont {Goold}, \citenamefont {Vinjanampathy},\ and\ \citenamefont {Modi}}]{Campaiolo2017}%
  \BibitemOpen
  \bibfield  {author} {\bibinfo {author} {\bibfnamefont {F.}~\bibnamefont {Campaioli}}, \bibinfo {author} {\bibfnamefont {F.~A.}\ \bibnamefont {Pollock}}, \bibinfo {author} {\bibfnamefont {F.~C.}\ \bibnamefont {Binder}}, \bibinfo {author} {\bibfnamefont {L.}~\bibnamefont {C\'eleri}}, \bibinfo {author} {\bibfnamefont {J.}~\bibnamefont {Goold}}, \bibinfo {author} {\bibfnamefont {S.}~\bibnamefont {Vinjanampathy}},\ and\ \bibinfo {author} {\bibfnamefont {K.}~\bibnamefont {Modi}},\ }\bibfield  {title} {\bibinfo {title} {Enhancing the charging power of quantum batteries},\ }\href {https://doi.org/10.1103/PhysRevLett.118.150601} {\bibfield  {journal} {\bibinfo  {journal} {Physical Review Letters}\ }\textbf {\bibinfo {volume} {118}},\ \bibinfo {pages} {150601} (\bibinfo {year} {2017})}\BibitemShut {NoStop}%
\bibitem [{\citenamefont {Ferraro}\ \emph {et~al.}(2018)\citenamefont {Ferraro}, \citenamefont {Campisi}, \citenamefont {Andolina}, \citenamefont {Pellegrini},\ and\ \citenamefont {Polini}}]{Ferraro2018}%
  \BibitemOpen
  \bibfield  {author} {\bibinfo {author} {\bibfnamefont {D.}~\bibnamefont {Ferraro}}, \bibinfo {author} {\bibfnamefont {M.}~\bibnamefont {Campisi}}, \bibinfo {author} {\bibfnamefont {G.~M.}\ \bibnamefont {Andolina}}, \bibinfo {author} {\bibfnamefont {V.}~\bibnamefont {Pellegrini}},\ and\ \bibinfo {author} {\bibfnamefont {M.}~\bibnamefont {Polini}},\ }\bibfield  {title} {\bibinfo {title} {High-power collective charging of a solid-state quantum battery},\ }\href {https://doi.org/10.1103/PhysRevLett.120.117702} {\bibfield  {journal} {\bibinfo  {journal} {Physical Review Letters}\ }\textbf {\bibinfo {volume} {120}},\ \bibinfo {pages} {117702} (\bibinfo {year} {2018})}\BibitemShut {NoStop}%
\bibitem [{\citenamefont {Juli\`a-Farr\'e}\ \emph {et~al.}(2020)\citenamefont {Juli\`a-Farr\'e}, \citenamefont {Salamon}, \citenamefont {Riera}, \citenamefont {Bera},\ and\ \citenamefont {Lewenstein}}]{Sergi2020}%
  \BibitemOpen
  \bibfield  {author} {\bibinfo {author} {\bibfnamefont {S.}~\bibnamefont {Juli\`a-Farr\'e}}, \bibinfo {author} {\bibfnamefont {T.}~\bibnamefont {Salamon}}, \bibinfo {author} {\bibfnamefont {A.}~\bibnamefont {Riera}}, \bibinfo {author} {\bibfnamefont {M.~N.}\ \bibnamefont {Bera}},\ and\ \bibinfo {author} {\bibfnamefont {M.}~\bibnamefont {Lewenstein}},\ }\bibfield  {title} {\bibinfo {title} {Bounds on the capacity and power of quantum batteries},\ }\href {https://doi.org/10.1103/PhysRevResearch.2.023113} {\bibfield  {journal} {\bibinfo  {journal} {Physical Review Research}\ }\textbf {\bibinfo {volume} {2}},\ \bibinfo {pages} {023113} (\bibinfo {year} {2020})}\BibitemShut {NoStop}%
\bibitem [{\citenamefont {Mayo}\ and\ \citenamefont {Roncaglia}(2022)}]{Mayo2022}%
  \BibitemOpen
  \bibfield  {author} {\bibinfo {author} {\bibfnamefont {F.}~\bibnamefont {Mayo}}\ and\ \bibinfo {author} {\bibfnamefont {A.~J.}\ \bibnamefont {Roncaglia}},\ }\bibfield  {title} {\bibinfo {title} {Collective effects and quantum coherence in dissipative charging of quantum batteries},\ }\href {https://doi.org/10.1103/PhysRevA.105.062203} {\bibfield  {journal} {\bibinfo  {journal} {Physical Review A}\ }\textbf {\bibinfo {volume} {105}},\ \bibinfo {pages} {062203} (\bibinfo {year} {2022})}\BibitemShut {NoStop}%
\bibitem [{\citenamefont {Gyhm}\ \emph {et~al.}(2022)\citenamefont {Gyhm}, \citenamefont {\ifmmode~\check{S}\else \v{S}\fi{}afr\'anek},\ and\ \citenamefont {Rosa}}]{Gyhm2022}%
  \BibitemOpen
  \bibfield  {author} {\bibinfo {author} {\bibfnamefont {J.-Y.}\ \bibnamefont {Gyhm}}, \bibinfo {author} {\bibfnamefont {D.}~\bibnamefont {\ifmmode~\check{S}\else \v{S}\fi{}afr\'anek}},\ and\ \bibinfo {author} {\bibfnamefont {D.}~\bibnamefont {Rosa}},\ }\bibfield  {title} {\bibinfo {title} {Quantum charging advantage cannot be extensive without global operations},\ }\href {https://doi.org/10.1103/PhysRevLett.128.140501} {\bibfield  {journal} {\bibinfo  {journal} {Physical Review Letters}\ }\textbf {\bibinfo {volume} {128}},\ \bibinfo {pages} {140501} (\bibinfo {year} {2022})}\BibitemShut {NoStop}%
\bibitem [{\citenamefont {Sugimoto}\ \emph {et~al.}(2025)\citenamefont {Sugimoto}, \citenamefont {Sagawa},\ and\ \citenamefont {Hamazaki}}]{Sugimoto2025}%
  \BibitemOpen
  \bibfield  {author} {\bibinfo {author} {\bibfnamefont {S.}~\bibnamefont {Sugimoto}}, \bibinfo {author} {\bibfnamefont {T.}~\bibnamefont {Sagawa}},\ and\ \bibinfo {author} {\bibfnamefont {R.}~\bibnamefont {Hamazaki}},\ }\bibfield  {title} {\bibinfo {title} {Optimal work extraction from finite-time closed quantum dynamics},\ }\href {https://arxiv.org/abs/2508.20512} {\bibfield  {journal} {\bibinfo  {journal} {arXiv:2508.20512}\ } (\bibinfo {year} {2025})}\BibitemShut {NoStop}%
\bibitem [{\citenamefont {Binder}\ \emph {et~al.}(2015)\citenamefont {Binder}, \citenamefont {Vinjanampathy}, \citenamefont {Modi},\ and\ \citenamefont {Goold}}]{Binder2015}%
  \BibitemOpen
  \bibfield  {author} {\bibinfo {author} {\bibfnamefont {F.~C.}\ \bibnamefont {Binder}}, \bibinfo {author} {\bibfnamefont {S.}~\bibnamefont {Vinjanampathy}}, \bibinfo {author} {\bibfnamefont {K.}~\bibnamefont {Modi}},\ and\ \bibinfo {author} {\bibfnamefont {J.}~\bibnamefont {Goold}},\ }\bibfield  {title} {\bibinfo {title} {Quantacell: powerful charging of quantum batteries},\ }\href {https://doi.org/10.1088/1367-2630/17/7/075015} {\bibfield  {journal} {\bibinfo  {journal} {New Journal of Physics}\ }\textbf {\bibinfo {volume} {17}},\ \bibinfo {pages} {075015} (\bibinfo {year} {2015})}\BibitemShut {NoStop}%
\bibitem [{\citenamefont {Mohan}\ and\ \citenamefont {Pati}(2022)}]{Mohan2022}%
  \BibitemOpen
  \bibfield  {author} {\bibinfo {author} {\bibfnamefont {B.}~\bibnamefont {Mohan}}\ and\ \bibinfo {author} {\bibfnamefont {A.~K.}\ \bibnamefont {Pati}},\ }\bibfield  {title} {\bibinfo {title} {Quantum speed limits for observables},\ }\href {https://doi.org/10.1103/PhysRevA.106.042436} {\bibfield  {journal} {\bibinfo  {journal} {Physical Review A}\ }\textbf {\bibinfo {volume} {106}},\ \bibinfo {pages} {042436} (\bibinfo {year} {2022})}\BibitemShut {NoStop}%
\bibitem [{\citenamefont {Mazzoncini}\ \emph {et~al.}(2023)\citenamefont {Mazzoncini}, \citenamefont {Cavina}, \citenamefont {Andolina}, \citenamefont {Erdman},\ and\ \citenamefont {Giovannetti}}]{Mazzoncini2013}%
  \BibitemOpen
  \bibfield  {author} {\bibinfo {author} {\bibfnamefont {F.}~\bibnamefont {Mazzoncini}}, \bibinfo {author} {\bibfnamefont {V.}~\bibnamefont {Cavina}}, \bibinfo {author} {\bibfnamefont {G.~M.}\ \bibnamefont {Andolina}}, \bibinfo {author} {\bibfnamefont {P.~A.}\ \bibnamefont {Erdman}},\ and\ \bibinfo {author} {\bibfnamefont {V.}~\bibnamefont {Giovannetti}},\ }\bibfield  {title} {\bibinfo {title} {Optimal control methods for quantum batteries},\ }\href {https://doi.org/10.1103/PhysRevA.107.032218} {\bibfield  {journal} {\bibinfo  {journal} {Physical Review A}\ }\textbf {\bibinfo {volume} {107}},\ \bibinfo {pages} {032218} (\bibinfo {year} {2023})}\BibitemShut {NoStop}%
\bibitem [{\citenamefont {Gyhm}\ \emph {et~al.}(2024)\citenamefont {Gyhm}, \citenamefont {Rosa},\ and\ \citenamefont {\ifmmode~\check{S}\else \v{S}\fi{}afr\'anek}}]{Gyhm2024}%
  \BibitemOpen
  \bibfield  {author} {\bibinfo {author} {\bibfnamefont {J.-Y.}\ \bibnamefont {Gyhm}}, \bibinfo {author} {\bibfnamefont {D.}~\bibnamefont {Rosa}},\ and\ \bibinfo {author} {\bibfnamefont {D.}~\bibnamefont {\ifmmode~\check{S}\else \v{S}\fi{}afr\'anek}},\ }\bibfield  {title} {\bibinfo {title} {Minimal time required to charge a quantum system},\ }\href {https://doi.org/10.1103/PhysRevA.109.022607} {\bibfield  {journal} {\bibinfo  {journal} {Physical Review A}\ }\textbf {\bibinfo {volume} {109}},\ \bibinfo {pages} {022607} (\bibinfo {year} {2024})}\BibitemShut {NoStop}%
\bibitem [{\citenamefont {Pati}\ \emph {et~al.}(2025)\citenamefont {Pati}, \citenamefont {Mohan}, \citenamefont {Sahil},\ and\ \citenamefont {Braunstein}}]{Mohan2025A}%
  \BibitemOpen
  \bibfield  {author} {\bibinfo {author} {\bibfnamefont {A.~K.}\ \bibnamefont {Pati}}, \bibinfo {author} {\bibfnamefont {B.}~\bibnamefont {Mohan}}, \bibinfo {author} {\bibnamefont {Sahil}},\ and\ \bibinfo {author} {\bibfnamefont {S.~L.}\ \bibnamefont {Braunstein}},\ }\bibfield  {title} {\bibinfo {title} {Exact quantum speed limits},\ }\href {https://doi.org/10.1103/mmzc-fyr9} {\bibfield  {journal} {\bibinfo  {journal} {Physical Review A}\ }\textbf {\bibinfo {volume} {112}},\ \bibinfo {pages} {062209} (\bibinfo {year} {2025})}\BibitemShut {NoStop}%
\bibitem [{\citenamefont {Rosa}\ \emph {et~al.}(2020)\citenamefont {Rosa}, \citenamefont {Rossini}, \citenamefont {Andolina}, \citenamefont {Polini},\ and\ \citenamefont {Carrega}}]{Rosa2020}%
  \BibitemOpen
  \bibfield  {author} {\bibinfo {author} {\bibfnamefont {D.}~\bibnamefont {Rosa}}, \bibinfo {author} {\bibfnamefont {D.}~\bibnamefont {Rossini}}, \bibinfo {author} {\bibfnamefont {G.~M.}\ \bibnamefont {Andolina}}, \bibinfo {author} {\bibfnamefont {M.}~\bibnamefont {Polini}},\ and\ \bibinfo {author} {\bibfnamefont {M.}~\bibnamefont {Carrega}},\ }\bibfield  {title} {\bibinfo {title} {Ultra-stable charging of fast-scrambling syk quantum batteries},\ }\href {https://doi.org/10.1007/JHEP11(2020)067} {\bibfield  {journal} {\bibinfo  {journal} {Journal of High Energy Physics}\ }\textbf {\bibinfo {volume} {2020}},\ \bibinfo {pages} {67} (\bibinfo {year} {2020})}\BibitemShut {NoStop}%
\bibitem [{\citenamefont {Quach}\ and\ \citenamefont {Munro}(2020)}]{Quach2020}%
  \BibitemOpen
  \bibfield  {author} {\bibinfo {author} {\bibfnamefont {J.~Q.}\ \bibnamefont {Quach}}\ and\ \bibinfo {author} {\bibfnamefont {W.~J.}\ \bibnamefont {Munro}},\ }\bibfield  {title} {\bibinfo {title} {Using dark states to charge and stabilize open quantum batteries},\ }\href {https://doi.org/10.1103/PhysRevApplied.14.024092} {\bibfield  {journal} {\bibinfo  {journal} {Physical Review Applied}\ }\textbf {\bibinfo {volume} {14}},\ \bibinfo {pages} {024092} (\bibinfo {year} {2020})}\BibitemShut {NoStop}%
\bibitem [{\citenamefont {Mohan}\ and\ \citenamefont {Pati}(2021)}]{Mohan2021}%
  \BibitemOpen
  \bibfield  {author} {\bibinfo {author} {\bibfnamefont {B.}~\bibnamefont {Mohan}}\ and\ \bibinfo {author} {\bibfnamefont {A.~K.}\ \bibnamefont {Pati}},\ }\bibfield  {title} {\bibinfo {title} {Reverse quantum speed limit: How slowly a quantum battery can discharge},\ }\href {https://doi.org/10.1103/PhysRevA.104.042209} {\bibfield  {journal} {\bibinfo  {journal} {Physical Review A}\ }\textbf {\bibinfo {volume} {104}},\ \bibinfo {pages} {042209} (\bibinfo {year} {2021})}\BibitemShut {NoStop}%
\bibitem [{\citenamefont {Grazi}\ \emph {et~al.}(2025)\citenamefont {Grazi}, \citenamefont {Johannesson}, \citenamefont {Ferraro},\ and\ \citenamefont {Ziani}}]{Grazi2025}%
  \BibitemOpen
  \bibfield  {author} {\bibinfo {author} {\bibfnamefont {R.}~\bibnamefont {Grazi}}, \bibinfo {author} {\bibfnamefont {H.}~\bibnamefont {Johannesson}}, \bibinfo {author} {\bibfnamefont {D.}~\bibnamefont {Ferraro}},\ and\ \bibinfo {author} {\bibfnamefont {N.~T.}\ \bibnamefont {Ziani}},\ }\bibfield  {title} {\bibinfo {title} {Finite-time protocols stabilize charging in noisy ising quantum batteries},\ }\href {https://arxiv.org/abs/2512.14521} {\bibfield  {journal} {\bibinfo  {journal} {arXiv:2512.14521}\ } (\bibinfo {year} {2025})}\BibitemShut {NoStop}%
\bibitem [{\citenamefont {Lu}\ \emph {et~al.}(2025)\citenamefont {Lu}, \citenamefont {Tian}, \citenamefont {L\"u},\ and\ \citenamefont {Shang}}]{Lu2025}%
  \BibitemOpen
  \bibfield  {author} {\bibinfo {author} {\bibfnamefont {Z.-G.}\ \bibnamefont {Lu}}, \bibinfo {author} {\bibfnamefont {G.}~\bibnamefont {Tian}}, \bibinfo {author} {\bibfnamefont {X.-Y.}\ \bibnamefont {L\"u}},\ and\ \bibinfo {author} {\bibfnamefont {C.}~\bibnamefont {Shang}},\ }\bibfield  {title} {\bibinfo {title} {Topological quantum batteries},\ }\href {https://doi.org/10.1103/PhysRevLett.134.180401} {\bibfield  {journal} {\bibinfo  {journal} {Physical Review Letters}\ }\textbf {\bibinfo {volume} {134}},\ \bibinfo {pages} {180401} (\bibinfo {year} {2025})}\BibitemShut {NoStop}%
\bibitem [{\citenamefont {Rossini}\ \emph {et~al.}(2020)\citenamefont {Rossini}, \citenamefont {Andolina}, \citenamefont {Rosa}, \citenamefont {Carrega},\ and\ \citenamefont {Polini}}]{Rossini2020}%
  \BibitemOpen
  \bibfield  {author} {\bibinfo {author} {\bibfnamefont {D.}~\bibnamefont {Rossini}}, \bibinfo {author} {\bibfnamefont {G.~M.}\ \bibnamefont {Andolina}}, \bibinfo {author} {\bibfnamefont {D.}~\bibnamefont {Rosa}}, \bibinfo {author} {\bibfnamefont {M.}~\bibnamefont {Carrega}},\ and\ \bibinfo {author} {\bibfnamefont {M.}~\bibnamefont {Polini}},\ }\bibfield  {title} {\bibinfo {title} {Quantum advantage in the charging process of sachdev-ye-kitaev batteries},\ }\href {https://doi.org/10.1103/PhysRevLett.125.236402} {\bibfield  {journal} {\bibinfo  {journal} {Physical Review Letters}\ }\textbf {\bibinfo {volume} {125}},\ \bibinfo {pages} {236402} (\bibinfo {year} {2020})}\BibitemShut {NoStop}%
\bibitem [{\citenamefont {Konar}\ \emph {et~al.}(2022)\citenamefont {Konar}, \citenamefont {Lakkaraju}, \citenamefont {Ghosh},\ and\ \citenamefont {Sen(De)}}]{Konar2022}%
  \BibitemOpen
  \bibfield  {author} {\bibinfo {author} {\bibfnamefont {T.~K.}\ \bibnamefont {Konar}}, \bibinfo {author} {\bibfnamefont {L.~G.~C.}\ \bibnamefont {Lakkaraju}}, \bibinfo {author} {\bibfnamefont {S.}~\bibnamefont {Ghosh}},\ and\ \bibinfo {author} {\bibfnamefont {A.}~\bibnamefont {Sen(De)}},\ }\bibfield  {title} {\bibinfo {title} {Quantum battery with ultracold atoms: Bosons versus fermions},\ }\href {https://doi.org/10.1103/PhysRevA.106.022618} {\bibfield  {journal} {\bibinfo  {journal} {Physical Review A}\ }\textbf {\bibinfo {volume} {106}},\ \bibinfo {pages} {022618} (\bibinfo {year} {2022})}\BibitemShut {NoStop}%
\bibitem [{\citenamefont {Shaghaghi}\ \emph {et~al.}(2022)\citenamefont {Shaghaghi}, \citenamefont {Singh}, \citenamefont {Benenti},\ and\ \citenamefont {Rosa}}]{Shaghaghi2022}%
  \BibitemOpen
  \bibfield  {author} {\bibinfo {author} {\bibfnamefont {V.}~\bibnamefont {Shaghaghi}}, \bibinfo {author} {\bibfnamefont {V.}~\bibnamefont {Singh}}, \bibinfo {author} {\bibfnamefont {G.}~\bibnamefont {Benenti}},\ and\ \bibinfo {author} {\bibfnamefont {D.}~\bibnamefont {Rosa}},\ }\bibfield  {title} {\bibinfo {title} {Micromasers as quantum batteries},\ }\href {https://doi.org/10.1088/2058-9565/ac8829} {\bibfield  {journal} {\bibinfo  {journal} {Quantum Science and Technology}\ }\textbf {\bibinfo {volume} {7}},\ \bibinfo {pages} {04LT01} (\bibinfo {year} {2022})}\BibitemShut {NoStop}%
\bibitem [{\citenamefont {Shukla}\ \emph {et~al.}(2025)\citenamefont {Shukla}, \citenamefont {Kumar}, \citenamefont {Sen},\ and\ \citenamefont {Mishra}}]{Shukla2025}%
  \BibitemOpen
  \bibfield  {author} {\bibinfo {author} {\bibfnamefont {R.~K.}\ \bibnamefont {Shukla}}, \bibinfo {author} {\bibfnamefont {R.}~\bibnamefont {Kumar}}, \bibinfo {author} {\bibfnamefont {U.}~\bibnamefont {Sen}},\ and\ \bibinfo {author} {\bibfnamefont {S.~K.}\ \bibnamefont {Mishra}},\ }\bibfield  {title} {\bibinfo {title} {Optimizing quantum battery performance by reducing battery influence in charging dynamics},\ }\href {https://arxiv.org/abs/2505.08029} {\bibfield  {journal} {\bibinfo  {journal} {arXiv:2505.08029}\ } (\bibinfo {year} {2025})}\BibitemShut {NoStop}%
\bibitem [{\citenamefont {Malavazi}\ \emph {et~al.}(2026)\citenamefont {Malavazi}, \citenamefont {Ahmadi}, \citenamefont {Horodecki},\ and\ \citenamefont {Dieguez}}]{Malavazi2026}%
  \BibitemOpen
  \bibfield  {author} {\bibinfo {author} {\bibfnamefont {A.~H.~A.}\ \bibnamefont {Malavazi}}, \bibinfo {author} {\bibfnamefont {B.}~\bibnamefont {Ahmadi}}, \bibinfo {author} {\bibfnamefont {P.}~\bibnamefont {Horodecki}},\ and\ \bibinfo {author} {\bibfnamefont {P.~R.}\ \bibnamefont {Dieguez}},\ }\bibfield  {title} {\bibinfo {title} {Charge-preserving operations in quantum batteries},\ }\href {https://doi.org/10.1103/2jtp-jpkn} {\bibfield  {journal} {\bibinfo  {journal} {PRX Energy}\ }\textbf {\bibinfo {volume} {5}},\ \bibinfo {pages} {023004} (\bibinfo {year} {2026})}\BibitemShut {NoStop}%
\bibitem [{\citenamefont {Ahmadi}\ \emph {et~al.}(2026)\citenamefont {Ahmadi}, \citenamefont {Malavazi}, \citenamefont {Splettstoesser}, \citenamefont {Horodecki},\ and\ \citenamefont {Du}}]{Ahmadi2026}%
  \BibitemOpen
  \bibfield  {author} {\bibinfo {author} {\bibfnamefont {B.}~\bibnamefont {Ahmadi}}, \bibinfo {author} {\bibfnamefont {A.~H.~A.}\ \bibnamefont {Malavazi}}, \bibinfo {author} {\bibfnamefont {J.}~\bibnamefont {Splettstoesser}}, \bibinfo {author} {\bibfnamefont {P.}~\bibnamefont {Horodecki}},\ and\ \bibinfo {author} {\bibfnamefont {L.}~\bibnamefont {Du}},\ }\bibfield  {title} {\bibinfo {title} {Charging quantum batteries with chiral squeezing},\ }\href {https://arxiv.org/abs/2606.16764} {\bibfield  {journal} {\bibinfo  {journal} {arXiv:2606.16764}\ } (\bibinfo {year} {2026})}\BibitemShut {NoStop}%
\bibitem [{\citenamefont {Joshi}\ and\ \citenamefont {Mahesh}(2022)}]{Joshi2022}%
  \BibitemOpen
  \bibfield  {author} {\bibinfo {author} {\bibfnamefont {J.}~\bibnamefont {Joshi}}\ and\ \bibinfo {author} {\bibfnamefont {T.~S.}\ \bibnamefont {Mahesh}},\ }\bibfield  {title} {\bibinfo {title} {Experimental investigation of a quantum battery using star-topology nmr spin systems},\ }\href {https://doi.org/10.1103/PhysRevA.106.042601} {\bibfield  {journal} {\bibinfo  {journal} {Physical Review A}\ }\textbf {\bibinfo {volume} {106}},\ \bibinfo {pages} {042601} (\bibinfo {year} {2022})}\BibitemShut {NoStop}%
\bibitem [{\citenamefont {Quach}\ \emph {et~al.}(2022)\citenamefont {Quach}, \citenamefont {McGhee}, \citenamefont {Ganzer}, \citenamefont {Rouse}, \citenamefont {Lovett}, \citenamefont {Gauger}, \citenamefont {Keeling}, \citenamefont {Cerullo}, \citenamefont {Lidzey},\ and\ \citenamefont {Virgili}}]{Quach2022}%
  \BibitemOpen
  \bibfield  {author} {\bibinfo {author} {\bibfnamefont {J.~Q.}\ \bibnamefont {Quach}}, \bibinfo {author} {\bibfnamefont {K.~E.}\ \bibnamefont {McGhee}}, \bibinfo {author} {\bibfnamefont {L.}~\bibnamefont {Ganzer}}, \bibinfo {author} {\bibfnamefont {D.~M.}\ \bibnamefont {Rouse}}, \bibinfo {author} {\bibfnamefont {B.~W.}\ \bibnamefont {Lovett}}, \bibinfo {author} {\bibfnamefont {E.~M.}\ \bibnamefont {Gauger}}, \bibinfo {author} {\bibfnamefont {J.}~\bibnamefont {Keeling}}, \bibinfo {author} {\bibfnamefont {G.}~\bibnamefont {Cerullo}}, \bibinfo {author} {\bibfnamefont {D.~G.}\ \bibnamefont {Lidzey}},\ and\ \bibinfo {author} {\bibfnamefont {T.}~\bibnamefont {Virgili}},\ }\bibfield  {title} {\bibinfo {title} {Superabsorption in an organic microcavity: Toward a quantum battery},\ }\href {https://doi.org/10.1126/sciadv.abk3160} {\bibfield  {journal} {\bibinfo  {journal} {Science Advances}\ }\textbf {\bibinfo {volume} {8}},\ \bibinfo {pages} {eabk3160} (\bibinfo {year} {2022})}\BibitemShut {NoStop}%
\bibitem [{\citenamefont {Huang}\ \emph {et~al.}(2023)\citenamefont {Huang}, \citenamefont {Wang}, \citenamefont {Xiao}, \citenamefont {Gao}, \citenamefont {Lin},\ and\ \citenamefont {Xue}}]{Huang2023}%
  \BibitemOpen
  \bibfield  {author} {\bibinfo {author} {\bibfnamefont {X.}~\bibnamefont {Huang}}, \bibinfo {author} {\bibfnamefont {K.}~\bibnamefont {Wang}}, \bibinfo {author} {\bibfnamefont {L.}~\bibnamefont {Xiao}}, \bibinfo {author} {\bibfnamefont {L.}~\bibnamefont {Gao}}, \bibinfo {author} {\bibfnamefont {H.}~\bibnamefont {Lin}},\ and\ \bibinfo {author} {\bibfnamefont {P.}~\bibnamefont {Xue}},\ }\bibfield  {title} {\bibinfo {title} {Demonstration of the charging progress of quantum batteries},\ }\href {https://doi.org/10.1103/PhysRevA.107.L030201} {\bibfield  {journal} {\bibinfo  {journal} {Physical Review A}\ }\textbf {\bibinfo {volume} {107}},\ \bibinfo {pages} {L030201} (\bibinfo {year} {2023})}\BibitemShut {NoStop}%
\bibitem [{\citenamefont {Camposeo}\ \emph {et~al.}(2025)\citenamefont {Camposeo}, \citenamefont {Virgili}, \citenamefont {Lombardi}, \citenamefont {Cerullo}, \citenamefont {Pisignano},\ and\ \citenamefont {Polini}}]{Camposeo2025}%
  \BibitemOpen
  \bibfield  {author} {\bibinfo {author} {\bibfnamefont {A.}~\bibnamefont {Camposeo}}, \bibinfo {author} {\bibfnamefont {T.}~\bibnamefont {Virgili}}, \bibinfo {author} {\bibfnamefont {F.}~\bibnamefont {Lombardi}}, \bibinfo {author} {\bibfnamefont {G.}~\bibnamefont {Cerullo}}, \bibinfo {author} {\bibfnamefont {D.}~\bibnamefont {Pisignano}},\ and\ \bibinfo {author} {\bibfnamefont {M.}~\bibnamefont {Polini}},\ }\bibfield  {title} {\bibinfo {title} {Quantum batteries: A materials science perspective},\ }\href {https://doi.org/10.1002/adma.202415073} {\bibfield  {journal} {\bibinfo  {journal} {Advanced Materials}\ }\textbf {\bibinfo {volume} {37}},\ \bibinfo {pages} {2415073} (\bibinfo {year} {2025})}\BibitemShut {NoStop}%
\bibitem [{\citenamefont {Tibben}\ \emph {et~al.}(2025)\citenamefont {Tibben}, \citenamefont {Della~Gaspera}, \citenamefont {van Embden}, \citenamefont {Reineck}, \citenamefont {Quach}, \citenamefont {Campaioli},\ and\ \citenamefont {G\'omez}}]{Tibben2025}%
  \BibitemOpen
  \bibfield  {author} {\bibinfo {author} {\bibfnamefont {D.~J.}\ \bibnamefont {Tibben}}, \bibinfo {author} {\bibfnamefont {E.}~\bibnamefont {Della~Gaspera}}, \bibinfo {author} {\bibfnamefont {J.}~\bibnamefont {van Embden}}, \bibinfo {author} {\bibfnamefont {P.}~\bibnamefont {Reineck}}, \bibinfo {author} {\bibfnamefont {J.~Q.}\ \bibnamefont {Quach}}, \bibinfo {author} {\bibfnamefont {F.}~\bibnamefont {Campaioli}},\ and\ \bibinfo {author} {\bibfnamefont {D.~E.}\ \bibnamefont {G\'omez}},\ }\bibfield  {title} {\bibinfo {title} {Extending the self-discharge time of dicke quantum batteries using molecular triplets},\ }\href {https://doi.org/10.1103/bhyh-53np} {\bibfield  {journal} {\bibinfo  {journal} {PRX Energy}\ }\textbf {\bibinfo {volume} {4}},\ \bibinfo {pages} {023012} (\bibinfo {year} {2025})}\BibitemShut {NoStop}%
\bibitem [{\citenamefont {Hu}\ \emph {et~al.}(2026)\citenamefont {Hu}, \citenamefont {Liu}, \citenamefont {Zhao}, \citenamefont {Zhong}, \citenamefont {Zhou}, \citenamefont {Liu}, \citenamefont {Yuan}, \citenamefont {Lin}, \citenamefont {Xu}, \citenamefont {Hu}, \citenamefont {Xie}, \citenamefont {Liu}, \citenamefont {Zhou}, \citenamefont {Ri}, \citenamefont {Zhang}, \citenamefont {Deng}, \citenamefont {Saguia}, \citenamefont {Linpeng}, \citenamefont {Sarandy}, \citenamefont {Liu}, \citenamefont {Santos}, \citenamefont {Tan},\ and\ \citenamefont {Yu}}]{Hu2026}%
  \BibitemOpen
  \bibfield  {author} {\bibinfo {author} {\bibfnamefont {C.-K.}\ \bibnamefont {Hu}}, \bibinfo {author} {\bibfnamefont {C.}~\bibnamefont {Liu}}, \bibinfo {author} {\bibfnamefont {J.}~\bibnamefont {Zhao}}, \bibinfo {author} {\bibfnamefont {L.}~\bibnamefont {Zhong}}, \bibinfo {author} {\bibfnamefont {Y.}~\bibnamefont {Zhou}}, \bibinfo {author} {\bibfnamefont {M.}~\bibnamefont {Liu}}, \bibinfo {author} {\bibfnamefont {H.}~\bibnamefont {Yuan}}, \bibinfo {author} {\bibfnamefont {Y.}~\bibnamefont {Lin}}, \bibinfo {author} {\bibfnamefont {Y.}~\bibnamefont {Xu}}, \bibinfo {author} {\bibfnamefont {G.}~\bibnamefont {Hu}}, \bibinfo {author} {\bibfnamefont {G.}~\bibnamefont {Xie}}, \bibinfo {author} {\bibfnamefont {Z.}~\bibnamefont {Liu}}, \bibinfo {author} {\bibfnamefont {R.}~\bibnamefont {Zhou}}, \bibinfo {author} {\bibfnamefont {Y.}~\bibnamefont {Ri}}, \bibinfo {author} {\bibfnamefont {W.}~\bibnamefont {Zhang}}, \bibinfo {author} {\bibfnamefont {R.}~\bibnamefont {Deng}}, \bibinfo {author} {\bibfnamefont
  {A.}~\bibnamefont {Saguia}}, \bibinfo {author} {\bibfnamefont {X.}~\bibnamefont {Linpeng}}, \bibinfo {author} {\bibfnamefont {M.~S.}\ \bibnamefont {Sarandy}}, \bibinfo {author} {\bibfnamefont {S.}~\bibnamefont {Liu}}, \bibinfo {author} {\bibfnamefont {A.~C.}\ \bibnamefont {Santos}}, \bibinfo {author} {\bibfnamefont {D.}~\bibnamefont {Tan}},\ and\ \bibinfo {author} {\bibfnamefont {D.}~\bibnamefont {Yu}},\ }\bibfield  {title} {\bibinfo {title} {Quantum charging advantage in superconducting solid-state batteries},\ }\href {https://doi.org/10.1103/sp5l-c6m8} {\bibfield  {journal} {\bibinfo  {journal} {Physical Review Letters}\ }\textbf {\bibinfo {volume} {136}},\ \bibinfo {pages} {060401} (\bibinfo {year} {2026})}\BibitemShut {NoStop}%
\bibitem [{\citenamefont {Campaioli}\ \emph {et~al.}(2024)\citenamefont {Campaioli}, \citenamefont {Gherardini}, \citenamefont {Quach}, \citenamefont {Polini},\ and\ \citenamefont {Andolina}}]{Campaioli2024}%
  \BibitemOpen
  \bibfield  {author} {\bibinfo {author} {\bibfnamefont {F.}~\bibnamefont {Campaioli}}, \bibinfo {author} {\bibfnamefont {S.}~\bibnamefont {Gherardini}}, \bibinfo {author} {\bibfnamefont {J.~Q.}\ \bibnamefont {Quach}}, \bibinfo {author} {\bibfnamefont {M.}~\bibnamefont {Polini}},\ and\ \bibinfo {author} {\bibfnamefont {G.~M.}\ \bibnamefont {Andolina}},\ }\bibfield  {title} {\bibinfo {title} {Colloquium: Quantum batteries},\ }\href {https://doi.org/10.1103/RevModPhys.96.031001} {\bibfield  {journal} {\bibinfo  {journal} {Reviews of Modern Physics}\ }\textbf {\bibinfo {volume} {96}},\ \bibinfo {pages} {031001} (\bibinfo {year} {2024})}\BibitemShut {NoStop}%
\bibitem [{\citenamefont {Friis}\ and\ \citenamefont {Huber}(2018)}]{Friis2018}%
  \BibitemOpen
  \bibfield  {author} {\bibinfo {author} {\bibfnamefont {N.}~\bibnamefont {Friis}}\ and\ \bibinfo {author} {\bibfnamefont {M.}~\bibnamefont {Huber}},\ }\bibfield  {title} {\bibinfo {title} {Precision and {W}ork {F}luctuations in {G}aussian {B}attery {C}harging},\ }\href {https://doi.org/10.22331/q-2018-04-23-61} {\bibfield  {journal} {\bibinfo  {journal} {{Quantum}}\ }\textbf {\bibinfo {volume} {2}},\ \bibinfo {pages} {61} (\bibinfo {year} {2018})}\BibitemShut {NoStop}%
\bibitem [{\citenamefont {McKay}\ \emph {et~al.}(2018)\citenamefont {McKay}, \citenamefont {Rodr\'{\i}guez-Briones},\ and\ \citenamefont {Mart\'{\i}n-Mart\'{\i}nez}}]{McKay2018}%
  \BibitemOpen
  \bibfield  {author} {\bibinfo {author} {\bibfnamefont {E.}~\bibnamefont {McKay}}, \bibinfo {author} {\bibfnamefont {N.~A.}\ \bibnamefont {Rodr\'{\i}guez-Briones}},\ and\ \bibinfo {author} {\bibfnamefont {E.}~\bibnamefont {Mart\'{\i}n-Mart\'{\i}nez}},\ }\bibfield  {title} {\bibinfo {title} {Fluctuations of work cost in optimal generation of correlations},\ }\href {https://doi.org/10.1103/PhysRevE.98.032132} {\bibfield  {journal} {\bibinfo  {journal} {Physical Review E}\ }\textbf {\bibinfo {volume} {98}},\ \bibinfo {pages} {032132} (\bibinfo {year} {2018})}\BibitemShut {NoStop}%
\bibitem [{\citenamefont {Perarnau-Llobet}\ and\ \citenamefont {Uzdin}(2019)}]{Llobet2019}%
  \BibitemOpen
  \bibfield  {author} {\bibinfo {author} {\bibfnamefont {M.}~\bibnamefont {Perarnau-Llobet}}\ and\ \bibinfo {author} {\bibfnamefont {R.}~\bibnamefont {Uzdin}},\ }\bibfield  {title} {\bibinfo {title} {Collective operations can extremely reduce work fluctuations},\ }\href {https://doi.org/10.1088/1367-2630/ab36a9} {\bibfield  {journal} {\bibinfo  {journal} {New Journal of Physics}\ }\textbf {\bibinfo {volume} {21}},\ \bibinfo {pages} {083023} (\bibinfo {year} {2019})}\BibitemShut {NoStop}%
\bibitem [{\citenamefont {Crescente}\ \emph {et~al.}(2020)\citenamefont {Crescente}, \citenamefont {Carrega}, \citenamefont {Sassetti},\ and\ \citenamefont {Ferraro}}]{Crescente2020}%
  \BibitemOpen
  \bibfield  {author} {\bibinfo {author} {\bibfnamefont {A.}~\bibnamefont {Crescente}}, \bibinfo {author} {\bibfnamefont {M.}~\bibnamefont {Carrega}}, \bibinfo {author} {\bibfnamefont {M.}~\bibnamefont {Sassetti}},\ and\ \bibinfo {author} {\bibfnamefont {D.}~\bibnamefont {Ferraro}},\ }\bibfield  {title} {\bibinfo {title} {Charging and energy fluctuations of a driven quantum battery},\ }\href {https://doi.org/10.1088/1367-2630/ab91fc} {\bibfield  {journal} {\bibinfo  {journal} {New Journal of Physics}\ }\textbf {\bibinfo {volume} {22}},\ \bibinfo {pages} {063057} (\bibinfo {year} {2020})}\BibitemShut {NoStop}%
\bibitem [{\citenamefont {Caravelli}\ \emph {et~al.}(2020)\citenamefont {Caravelli}, \citenamefont {Coulter-De~Wit}, \citenamefont {Garc\'{\i}a-Pintos},\ and\ \citenamefont {Hamma}}]{Caravelli2020}%
  \BibitemOpen
  \bibfield  {author} {\bibinfo {author} {\bibfnamefont {F.}~\bibnamefont {Caravelli}}, \bibinfo {author} {\bibfnamefont {G.}~\bibnamefont {Coulter-De~Wit}}, \bibinfo {author} {\bibfnamefont {L.~P.}\ \bibnamefont {Garc\'{\i}a-Pintos}},\ and\ \bibinfo {author} {\bibfnamefont {A.}~\bibnamefont {Hamma}},\ }\bibfield  {title} {\bibinfo {title} {Random quantum batteries},\ }\href {https://doi.org/10.1103/PhysRevResearch.2.023095} {\bibfield  {journal} {\bibinfo  {journal} {Physical Review Research}\ }\textbf {\bibinfo {volume} {2}},\ \bibinfo {pages} {023095} (\bibinfo {year} {2020})}\BibitemShut {NoStop}%
\bibitem [{\citenamefont {Garc\'{\i}a-Pintos}\ \emph {et~al.}(2020)\citenamefont {Garc\'{\i}a-Pintos}, \citenamefont {Hamma},\ and\ \citenamefont {del Campo}}]{Pintos2020}%
  \BibitemOpen
  \bibfield  {author} {\bibinfo {author} {\bibfnamefont {L.~P.}\ \bibnamefont {Garc\'{\i}a-Pintos}}, \bibinfo {author} {\bibfnamefont {A.}~\bibnamefont {Hamma}},\ and\ \bibinfo {author} {\bibfnamefont {A.}~\bibnamefont {del Campo}},\ }\bibfield  {title} {\bibinfo {title} {Fluctuations in extractable work bound the charging power of quantum batteries},\ }\href {https://doi.org/10.1103/PhysRevLett.125.040601} {\bibfield  {journal} {\bibinfo  {journal} {Physical Review Letters}\ }\textbf {\bibinfo {volume} {125}},\ \bibinfo {pages} {040601} (\bibinfo {year} {2020})}\BibitemShut {NoStop}%
\bibitem [{\citenamefont {Bakhshinezhad}\ \emph {et~al.}(2024)\citenamefont {Bakhshinezhad}, \citenamefont {Jablonski}, \citenamefont {Binder},\ and\ \citenamefont {Friis}}]{Bakhshinezhad2024}%
  \BibitemOpen
  \bibfield  {author} {\bibinfo {author} {\bibfnamefont {P.}~\bibnamefont {Bakhshinezhad}}, \bibinfo {author} {\bibfnamefont {B.~R.}\ \bibnamefont {Jablonski}}, \bibinfo {author} {\bibfnamefont {F.~C.}\ \bibnamefont {Binder}},\ and\ \bibinfo {author} {\bibfnamefont {N.}~\bibnamefont {Friis}},\ }\bibfield  {title} {\bibinfo {title} {Trade-offs between precision and fluctuations in charging finite-dimensional quantum batteries},\ }\href {https://doi.org/10.1103/PhysRevE.109.014131} {\bibfield  {journal} {\bibinfo  {journal} {Physical Review E}\ }\textbf {\bibinfo {volume} {109}},\ \bibinfo {pages} {014131} (\bibinfo {year} {2024})}\BibitemShut {NoStop}%
\bibitem [{\citenamefont {Imai}\ \emph {et~al.}(2023)\citenamefont {Imai}, \citenamefont {G\"uhne},\ and\ \citenamefont {Nimmrichter}}]{Imai2023}%
  \BibitemOpen
  \bibfield  {author} {\bibinfo {author} {\bibfnamefont {S.}~\bibnamefont {Imai}}, \bibinfo {author} {\bibfnamefont {O.}~\bibnamefont {G\"uhne}},\ and\ \bibinfo {author} {\bibfnamefont {S.}~\bibnamefont {Nimmrichter}},\ }\bibfield  {title} {\bibinfo {title} {Work fluctuations and entanglement in quantum batteries},\ }\href {https://doi.org/10.1103/PhysRevA.107.022215} {\bibfield  {journal} {\bibinfo  {journal} {Physical Review A}\ }\textbf {\bibinfo {volume} {107}},\ \bibinfo {pages} {022215} (\bibinfo {year} {2023})}\BibitemShut {NoStop}%
\bibitem [{\citenamefont {Sarkar}\ \emph {et~al.}(2025)\citenamefont {Sarkar}, \citenamefont {Chaki}, \citenamefont {Ghosh},\ and\ \citenamefont {Sen}}]{Sarkar2025}%
  \BibitemOpen
  \bibfield  {author} {\bibinfo {author} {\bibfnamefont {A.}~\bibnamefont {Sarkar}}, \bibinfo {author} {\bibfnamefont {P.}~\bibnamefont {Chaki}}, \bibinfo {author} {\bibfnamefont {P.}~\bibnamefont {Ghosh}},\ and\ \bibinfo {author} {\bibfnamefont {U.}~\bibnamefont {Sen}},\ }\bibfield  {title} {\bibinfo {title} {Fluctuation in energy extraction from quantum batteries: How open should the system be to control it?},\ }\href {https://arxiv.org/abs/2505.16851} {\bibfield  {journal} {\bibinfo  {journal} {arXiv:2505.16851}\ } (\bibinfo {year} {2025})}\BibitemShut {NoStop}%
\bibitem [{\citenamefont {Rinaldi}\ \emph {et~al.}(2025)\citenamefont {Rinaldi}, \citenamefont {Filip}, \citenamefont {Gerace},\ and\ \citenamefont {Guarnieri}}]{Rinaldi2025}%
  \BibitemOpen
  \bibfield  {author} {\bibinfo {author} {\bibfnamefont {D.}~\bibnamefont {Rinaldi}}, \bibinfo {author} {\bibfnamefont {R.}~\bibnamefont {Filip}}, \bibinfo {author} {\bibfnamefont {D.}~\bibnamefont {Gerace}},\ and\ \bibinfo {author} {\bibfnamefont {G.}~\bibnamefont {Guarnieri}},\ }\bibfield  {title} {\bibinfo {title} {Reliable quantum advantage in quantum battery charging},\ }\href {https://doi.org/10.1103/6kwv-z6fx} {\bibfield  {journal} {\bibinfo  {journal} {Physical Review A}\ }\textbf {\bibinfo {volume} {112}},\ \bibinfo {pages} {012205} (\bibinfo {year} {2025})}\BibitemShut {NoStop}%
\bibitem [{\citenamefont {Polo}\ and\ \citenamefont {Centrone}(2025)}]{Polo2025}%
  \BibitemOpen
  \bibfield  {author} {\bibinfo {author} {\bibfnamefont {B.}~\bibnamefont {Polo}}\ and\ \bibinfo {author} {\bibfnamefont {F.}~\bibnamefont {Centrone}},\ }\bibfield  {title} {\bibinfo {title} {Precision bounds for bosonic quantum batteries},\ }\href {https://arxiv.org/abs/2505.24604} {\bibfield  {journal} {\bibinfo  {journal} {arXiv:2505.24604}\ } (\bibinfo {year} {2025})}\BibitemShut {NoStop}%
\bibitem [{\citenamefont {Allahverdyan}\ \emph {et~al.}(2004)\citenamefont {Allahverdyan}, \citenamefont {Balian},\ and\ \citenamefont {Nieuwenhuizen}}]{Allahverdyan2004}%
  \BibitemOpen
  \bibfield  {author} {\bibinfo {author} {\bibfnamefont {A.~E.}\ \bibnamefont {Allahverdyan}}, \bibinfo {author} {\bibfnamefont {R.}~\bibnamefont {Balian}},\ and\ \bibinfo {author} {\bibfnamefont {T.~M.}\ \bibnamefont {Nieuwenhuizen}},\ }\bibfield  {title} {\bibinfo {title} {Maximal work extraction from finite quantum systems},\ }\href {https://doi.org/10.1209/epl/i2004-10101-2} {\bibfield  {journal} {\bibinfo  {journal} {Europhysics Letters}\ }\textbf {\bibinfo {volume} {67}},\ \bibinfo {pages} {565} (\bibinfo {year} {2004})}\BibitemShut {NoStop}%
\bibitem [{\citenamefont {Esposito}\ \emph {et~al.}(2009)\citenamefont {Esposito}, \citenamefont {Harbola},\ and\ \citenamefont {Mukamel}}]{Esposito2009}%
  \BibitemOpen
  \bibfield  {author} {\bibinfo {author} {\bibfnamefont {M.}~\bibnamefont {Esposito}}, \bibinfo {author} {\bibfnamefont {U.}~\bibnamefont {Harbola}},\ and\ \bibinfo {author} {\bibfnamefont {S.}~\bibnamefont {Mukamel}},\ }\bibfield  {title} {\bibinfo {title} {Nonequilibrium fluctuations, fluctuation theorems, and counting statistics in quantum systems},\ }\href {https://doi.org/10.1103/RevModPhys.81.1665} {\bibfield  {journal} {\bibinfo  {journal} {Reviews of Modern Physics}\ }\textbf {\bibinfo {volume} {81}},\ \bibinfo {pages} {1665} (\bibinfo {year} {2009})}\BibitemShut {NoStop}%
\bibitem [{\citenamefont {Allahverdyan}\ and\ \citenamefont {Nieuwenhuizen}(2005)}]{Allahverdyan2005}%
  \BibitemOpen
  \bibfield  {author} {\bibinfo {author} {\bibfnamefont {A.~E.}\ \bibnamefont {Allahverdyan}}\ and\ \bibinfo {author} {\bibfnamefont {T.~M.}\ \bibnamefont {Nieuwenhuizen}},\ }\bibfield  {title} {\bibinfo {title} {Fluctuations of work from quantum subensembles: The case against quantum work-fluctuation theorems},\ }\href {https://doi.org/10.1103/PhysRevE.71.066102} {\bibfield  {journal} {\bibinfo  {journal} {Physical Review E}\ }\textbf {\bibinfo {volume} {71}},\ \bibinfo {pages} {066102} (\bibinfo {year} {2005})}\BibitemShut {NoStop}%
\bibitem [{\citenamefont {Allahverdyan}(2014)}]{Allahverdyan2014}%
  \BibitemOpen
  \bibfield  {author} {\bibinfo {author} {\bibfnamefont {A.~E.}\ \bibnamefont {Allahverdyan}},\ }\bibfield  {title} {\bibinfo {title} {Nonequilibrium quantum fluctuations of work},\ }\href {https://doi.org/10.1103/PhysRevE.90.032137} {\bibfield  {journal} {\bibinfo  {journal} {Physical Review E}\ }\textbf {\bibinfo {volume} {90}},\ \bibinfo {pages} {032137} (\bibinfo {year} {2014})}\BibitemShut {NoStop}%
\bibitem [{\citenamefont {Solinas}\ and\ \citenamefont {Gasparinetti}(2015)}]{Solinas2015}%
  \BibitemOpen
  \bibfield  {author} {\bibinfo {author} {\bibfnamefont {P.}~\bibnamefont {Solinas}}\ and\ \bibinfo {author} {\bibfnamefont {S.}~\bibnamefont {Gasparinetti}},\ }\bibfield  {title} {\bibinfo {title} {Full distribution of work done on a quantum system for arbitrary initial states},\ }\href {https://doi.org/10.1103/PhysRevE.92.042150} {\bibfield  {journal} {\bibinfo  {journal} {Physical Review E}\ }\textbf {\bibinfo {volume} {92}},\ \bibinfo {pages} {042150} (\bibinfo {year} {2015})}\BibitemShut {NoStop}%
\bibitem [{\citenamefont {Talkner}\ and\ \citenamefont {H\"anggi}(2016)}]{Talkner2016}%
  \BibitemOpen
  \bibfield  {author} {\bibinfo {author} {\bibfnamefont {P.}~\bibnamefont {Talkner}}\ and\ \bibinfo {author} {\bibfnamefont {P.}~\bibnamefont {H\"anggi}},\ }\bibfield  {title} {\bibinfo {title} {Aspects of quantum work},\ }\href {https://doi.org/10.1103/PhysRevE.93.022131} {\bibfield  {journal} {\bibinfo  {journal} {Physical Review E}\ }\textbf {\bibinfo {volume} {93}},\ \bibinfo {pages} {022131} (\bibinfo {year} {2016})}\BibitemShut {NoStop}%
\bibitem [{\citenamefont {Solinas}\ and\ \citenamefont {Gasparinetti}(2016)}]{Solinas2016}%
  \BibitemOpen
  \bibfield  {author} {\bibinfo {author} {\bibfnamefont {P.}~\bibnamefont {Solinas}}\ and\ \bibinfo {author} {\bibfnamefont {S.}~\bibnamefont {Gasparinetti}},\ }\bibfield  {title} {\bibinfo {title} {Probing quantum interference effects in the work distribution},\ }\href {https://doi.org/10.1103/PhysRevA.94.052103} {\bibfield  {journal} {\bibinfo  {journal} {Physical Review A}\ }\textbf {\bibinfo {volume} {94}},\ \bibinfo {pages} {052103} (\bibinfo {year} {2016})}\BibitemShut {NoStop}%
\bibitem [{\citenamefont {Xu}\ \emph {et~al.}(2018)\citenamefont {Xu}, \citenamefont {Zou}, \citenamefont {Guo},\ and\ \citenamefont {Kong}}]{Xu2018}%
  \BibitemOpen
  \bibfield  {author} {\bibinfo {author} {\bibfnamefont {B.-M.}\ \bibnamefont {Xu}}, \bibinfo {author} {\bibfnamefont {J.}~\bibnamefont {Zou}}, \bibinfo {author} {\bibfnamefont {L.-S.}\ \bibnamefont {Guo}},\ and\ \bibinfo {author} {\bibfnamefont {X.-M.}\ \bibnamefont {Kong}},\ }\bibfield  {title} {\bibinfo {title} {Effects of quantum coherence on work statistics},\ }\href {https://doi.org/10.1103/PhysRevA.97.052122} {\bibfield  {journal} {\bibinfo  {journal} {Physical Review A}\ }\textbf {\bibinfo {volume} {97}},\ \bibinfo {pages} {052122} (\bibinfo {year} {2018})}\BibitemShut {NoStop}%
\bibitem [{\citenamefont {Francica}(2022)}]{Francica2022}%
  \BibitemOpen
  \bibfield  {author} {\bibinfo {author} {\bibfnamefont {G.}~\bibnamefont {Francica}},\ }\bibfield  {title} {\bibinfo {title} {Most general class of quasiprobability distributions of work},\ }\href {https://doi.org/10.1103/PhysRevE.106.054129} {\bibfield  {journal} {\bibinfo  {journal} {Physical Review E}\ }\textbf {\bibinfo {volume} {106}},\ \bibinfo {pages} {054129} (\bibinfo {year} {2022})}\BibitemShut {NoStop}%
\bibitem [{\citenamefont {Francica}\ and\ \citenamefont {Dell'Anna}(2023)}]{Francica2023}%
  \BibitemOpen
  \bibfield  {author} {\bibinfo {author} {\bibfnamefont {G.}~\bibnamefont {Francica}}\ and\ \bibinfo {author} {\bibfnamefont {L.}~\bibnamefont {Dell'Anna}},\ }\bibfield  {title} {\bibinfo {title} {Quasiprobability distribution of work in the quantum ising model},\ }\href {https://doi.org/10.1103/PhysRevE.108.014106} {\bibfield  {journal} {\bibinfo  {journal} {Physical Review E}\ }\textbf {\bibinfo {volume} {108}},\ \bibinfo {pages} {014106} (\bibinfo {year} {2023})}\BibitemShut {NoStop}%
\bibitem [{\citenamefont {Hern\'andez-G\'omez}\ \emph {et~al.}(2024)\citenamefont {Hern\'andez-G\'omez}, \citenamefont {Gherardini}, \citenamefont {Belenchia}, \citenamefont {Lostaglio}, \citenamefont {Levy},\ and\ \citenamefont {Fabbri}}]{Santiago2024}%
  \BibitemOpen
  \bibfield  {author} {\bibinfo {author} {\bibfnamefont {S.}~\bibnamefont {Hern\'andez-G\'omez}}, \bibinfo {author} {\bibfnamefont {S.}~\bibnamefont {Gherardini}}, \bibinfo {author} {\bibfnamefont {A.}~\bibnamefont {Belenchia}}, \bibinfo {author} {\bibfnamefont {M.}~\bibnamefont {Lostaglio}}, \bibinfo {author} {\bibfnamefont {A.}~\bibnamefont {Levy}},\ and\ \bibinfo {author} {\bibfnamefont {N.}~\bibnamefont {Fabbri}},\ }\bibfield  {title} {\bibinfo {title} {Projective measurements can probe nonclassical work extraction and time correlations},\ }\href {https://doi.org/10.1103/PhysRevResearch.6.023280} {\bibfield  {journal} {\bibinfo  {journal} {Physical Review Research}\ }\textbf {\bibinfo {volume} {6}},\ \bibinfo {pages} {023280} (\bibinfo {year} {2024})}\BibitemShut {NoStop}%
\bibitem [{\citenamefont {Talkner}\ \emph {et~al.}(2007)\citenamefont {Talkner}, \citenamefont {Lutz},\ and\ \citenamefont {H\"anggi}}]{Talkner2007}%
  \BibitemOpen
  \bibfield  {author} {\bibinfo {author} {\bibfnamefont {P.}~\bibnamefont {Talkner}}, \bibinfo {author} {\bibfnamefont {E.}~\bibnamefont {Lutz}},\ and\ \bibinfo {author} {\bibfnamefont {P.}~\bibnamefont {H\"anggi}},\ }\bibfield  {title} {\bibinfo {title} {Fluctuation theorems: Work is not an observable},\ }\href {https://doi.org/10.1103/PhysRevE.75.050102} {\bibfield  {journal} {\bibinfo  {journal} {Physical Review E}\ }\textbf {\bibinfo {volume} {75}},\ \bibinfo {pages} {050102(R)} (\bibinfo {year} {2007})}\BibitemShut {NoStop}%
\bibitem [{\citenamefont {Hovhannisyan}\ and\ \citenamefont {Imparato}(2024)}]{Hovhannisyan2024}%
  \BibitemOpen
  \bibfield  {author} {\bibinfo {author} {\bibfnamefont {K.~V.}\ \bibnamefont {Hovhannisyan}}\ and\ \bibinfo {author} {\bibfnamefont {A.}~\bibnamefont {Imparato}},\ }\bibfield  {title} {\bibinfo {title} {Energy conservation and fluctuation theorem are incompatible for quantum work},\ }\href {https://doi.org/10.22331/q-2024-05-06-1336} {\bibfield  {journal} {\bibinfo  {journal} {{Quantum}}\ }\textbf {\bibinfo {volume} {8}},\ \bibinfo {pages} {1336} (\bibinfo {year} {2024})}\BibitemShut {NoStop}%
\bibitem [{\citenamefont {Pinto~Silva}\ and\ \citenamefont {Gelbwaser-Klimovsky}(2024)}]{Silva2024}%
  \BibitemOpen
  \bibfield  {author} {\bibinfo {author} {\bibfnamefont {T.~A.~B.}\ \bibnamefont {Pinto~Silva}}\ and\ \bibinfo {author} {\bibfnamefont {D.}~\bibnamefont {Gelbwaser-Klimovsky}},\ }\bibfield  {title} {\bibinfo {title} {Quantum work: Reconciling quantum mechanics and thermodynamics},\ }\href {https://doi.org/10.1103/PhysRevResearch.6.L022036} {\bibfield  {journal} {\bibinfo  {journal} {Physical Review Research}\ }\textbf {\bibinfo {volume} {6}},\ \bibinfo {pages} {L022036} (\bibinfo {year} {2024})}\BibitemShut {NoStop}%
\bibitem [{\citenamefont {Rubino}\ \emph {et~al.}(2025)\citenamefont {Rubino}, \citenamefont {Hovhannisyan},\ and\ \citenamefont {Skrzypczyk}}]{Rubino2025}%
  \BibitemOpen
  \bibfield  {author} {\bibinfo {author} {\bibfnamefont {G.}~\bibnamefont {Rubino}}, \bibinfo {author} {\bibfnamefont {K.~V.}\ \bibnamefont {Hovhannisyan}},\ and\ \bibinfo {author} {\bibfnamefont {P.}~\bibnamefont {Skrzypczyk}},\ }\bibfield  {title} {\bibinfo {title} {Revising the quantum work fluctuation framework to encompass energy conservation},\ }\href {https://doi.org/10.1038/s41534-025-01053-6} {\bibfield  {journal} {\bibinfo  {journal} {npj Quantum Information}\ }\textbf {\bibinfo {volume} {11}},\ \bibinfo {pages} {102} (\bibinfo {year} {2025})}\BibitemShut {NoStop}%
\bibitem [{\citenamefont {Robertson}(1929)}]{Robertson1929}%
  \BibitemOpen
  \bibfield  {author} {\bibinfo {author} {\bibfnamefont {H.~P.}\ \bibnamefont {Robertson}},\ }\bibfield  {title} {\bibinfo {title} {The uncertainty principle},\ }\href {https://doi.org/10.1103/PhysRev.34.163} {\bibfield  {journal} {\bibinfo  {journal} {Physical Review}\ }\textbf {\bibinfo {volume} {34}},\ \bibinfo {pages} {163} (\bibinfo {year} {1929})}\BibitemShut {NoStop}%
\bibitem [{\citenamefont {Schr{\"o}dinger}(1930)}]{Schrodinger1930}%
  \BibitemOpen
  \bibfield  {author} {\bibinfo {author} {\bibfnamefont {E.}~\bibnamefont {Schr{\"o}dinger}},\ }\bibfield  {title} {\bibinfo {title} {Zum heisenbergschen unsch{\"a}rfeprinzip},\ }\href@noop {} {\bibfield  {journal} {\bibinfo  {journal} {Sitzungsberichte der Preussischen Akademie der Wissenschaften, Physikalisch-mathematische Klasse}\ }\textbf {\bibinfo {volume} {14}},\ \bibinfo {pages} {296} (\bibinfo {year} {1930})}\BibitemShut {NoStop}%
\bibitem [{\citenamefont {Gaines}(1967)}]{Gaines1967}%
  \BibitemOpen
  \bibfield  {author} {\bibinfo {author} {\bibfnamefont {F.}~\bibnamefont {Gaines}},\ }\bibfield  {title} {\bibinfo {title} {On the arithmetic mean-geometric mean inequality},\ }\href {https://doi.org/10.2307/2316036} {\bibfield  {journal} {\bibinfo  {journal} {American Mathematical Monthly}\ }\textbf {\bibinfo {volume} {74}},\ \bibinfo {pages} {305} (\bibinfo {year} {1967})}\BibitemShut {NoStop}%
\bibitem [{\citenamefont {Maccone}\ and\ \citenamefont {Pati}(2014)}]{Maccone2014}%
  \BibitemOpen
  \bibfield  {author} {\bibinfo {author} {\bibfnamefont {L.}~\bibnamefont {Maccone}}\ and\ \bibinfo {author} {\bibfnamefont {A.~K.}\ \bibnamefont {Pati}},\ }\bibfield  {title} {\bibinfo {title} {Stronger uncertainty relations for all incompatible observables},\ }\href {https://doi.org/10.1103/PhysRevLett.113.260401} {\bibfield  {journal} {\bibinfo  {journal} {Physical Review Letters}\ }\textbf {\bibinfo {volume} {113}},\ \bibinfo {pages} {260401} (\bibinfo {year} {2014})}\BibitemShut {NoStop}%
\bibitem [{\citenamefont {Sathe}\ \emph {et~al.}(2024)\citenamefont {Sathe}, \citenamefont {García-Pintos},\ and\ \citenamefont {Caravelli}}]{Sathe2024}%
  \BibitemOpen
  \bibfield  {author} {\bibinfo {author} {\bibfnamefont {P.}~\bibnamefont {Sathe}}, \bibinfo {author} {\bibfnamefont {L.~P.}\ \bibnamefont {García-Pintos}},\ and\ \bibinfo {author} {\bibfnamefont {F.}~\bibnamefont {Caravelli}},\ }\bibfield  {title} {\bibinfo {title} {Operator-based quantum thermodynamic uncertainty relations},\ }\href {https://arxiv.org/abs/2406.11974} {\bibfield  {journal} {\bibinfo  {journal} {arXiv:2406.11974}\ } (\bibinfo {year} {2024})}\BibitemShut {NoStop}%
\bibitem [{\citenamefont {Gyhm}\ and\ \citenamefont {Fischer}(2024)}]{Gyhm20240}%
  \BibitemOpen
  \bibfield  {author} {\bibinfo {author} {\bibfnamefont {J.-Y.}\ \bibnamefont {Gyhm}}\ and\ \bibinfo {author} {\bibfnamefont {U.~R.}\ \bibnamefont {Fischer}},\ }\bibfield  {title} {\bibinfo {title} {Beneficial and detrimental entanglement for quantum battery charging},\ }\href {https://doi.org/10.1116/5.0184903} {\bibfield  {journal} {\bibinfo  {journal} {AVS Quantum Science}\ }\textbf {\bibinfo {volume} {6}},\ \bibinfo {pages} {012001} (\bibinfo {year} {2024})}\BibitemShut {NoStop}%
\bibitem [{\citenamefont {Shukla}\ \emph {et~al.}(2026)\citenamefont {Shukla}, \citenamefont {Mishra},\ and\ \citenamefont {Sen}}]{Shukla2026}%
  \BibitemOpen
  \bibfield  {author} {\bibinfo {author} {\bibfnamefont {R.~K.}\ \bibnamefont {Shukla}}, \bibinfo {author} {\bibfnamefont {S.~K.}\ \bibnamefont {Mishra}},\ and\ \bibinfo {author} {\bibfnamefont {U.}~\bibnamefont {Sen}},\ }\bibfield  {title} {\bibinfo {title} {Collective dynamics versus entanglement in quantum battery performance},\ }\href {https://arxiv.org/abs/2601.03119} {\bibfield  {journal} {\bibinfo  {journal} {arXiv:2601.03119}\ } (\bibinfo {year} {2026})}\BibitemShut {NoStop}%
\bibitem [{\citenamefont {for Disarmament}(2026)}]{Disarmament2026}%
  \BibitemOpen
  \bibfield  {author} {\bibinfo {author} {\bibfnamefont {Q.~S.}\ \bibnamefont {for Disarmament}},\ }\bibfield  {title} {\bibinfo {title} {Quantum scientists for disarmament: a manifesto},\ }\href {https://arxiv.org/abs/2601.14282} {\bibfield  {journal} {\bibinfo  {journal} {arXiv2601.14282}\ } (\bibinfo {year} {2026})}\BibitemShut {NoStop}%
\bibitem [{\citenamefont {Nishiyama}(2021)}]{Nishiyama2021}%
  \BibitemOpen
  \bibfield  {author} {\bibinfo {author} {\bibfnamefont {T.}~\bibnamefont {Nishiyama}},\ }\bibfield  {title} {\bibinfo {title} {A tight lower bound for the {Hellinger} distance with given means and variances},\ }\href {https://arxiv.org/abs/2010.13548} {\bibfield  {journal} {\bibinfo  {journal} {arXiv:2010.13548}\ } (\bibinfo {year} {2021})}\BibitemShut {NoStop}%
\bibitem [{\citenamefont {Nishiyama}\ and\ \citenamefont {Hasegawa}(2025)}]{Nishiyama2025}%
  \BibitemOpen
  \bibfield  {author} {\bibinfo {author} {\bibfnamefont {T.}~\bibnamefont {Nishiyama}}\ and\ \bibinfo {author} {\bibfnamefont {Y.}~\bibnamefont {Hasegawa}},\ }\bibfield  {title} {\bibinfo {title} {Speed limits and thermodynamic uncertainty relations for quantum systems with the non-{Hermitian} {Hamiltonian}},\ }\href {https://doi.org/10.1103/PhysRevA.111.012214} {\bibfield  {journal} {\bibinfo  {journal} {Physical Review A}\ }\textbf {\bibinfo {volume} {111}},\ \bibinfo {pages} {012214} (\bibinfo {year} {2025})}\BibitemShut {NoStop}%
\bibitem [{\citenamefont {Hasegawa}(2023)}]{Hasegawa2023}%
  \BibitemOpen
  \bibfield  {author} {\bibinfo {author} {\bibfnamefont {Y.}~\bibnamefont {Hasegawa}},\ }\bibfield  {title} {\bibinfo {title} {Unifying speed limit, thermodynamic uncertainty relation and {Heisenberg} principle via bulk-boundary correspondence},\ }\href {https://doi.org/10.1038/s41467-023-38074-8} {\bibfield  {journal} {\bibinfo  {journal} {Nature Communications}\ }\textbf {\bibinfo {volume} {14}},\ \bibinfo {pages} {2828} (\bibinfo {year} {2023})}\BibitemShut {NoStop}%
\end{thebibliography}%

\end{document}